\documentclass[letterpaper]{quantumarticle}
\pdfoutput=1

\usepackage{tikz}
\usetikzlibrary{decorations,decorations.pathmorphing,decorations.markings,calc,snakes,positioning,fixedpointarithmetic,shapes,shapes.geometric,patterns}

\usepackage{etoolbox}
\AtBeginEnvironment{tikzpicture}{\catcode`$=3 } 

\tikzset{alignmid/.style={baseline={([yshift=-.5ex]current bounding box.center)}}} 
\tikzset{every picture/.append style=alignmid}

\tikzset{
bottomzigzag/.style={postaction={draw,decorate, decoration={zigzag,amplitude=1pt,segment length=3pt,raise=1pt}}},
zigzag/.style={draw,decorate, decoration={zigzag,amplitude=1pt,segment length=3pt}},
rc/.style=rounded corners,
}

\tikzset{
    -|/.style={to path={-| (\tikztotarget)}},
    |-/.style={to path={|- (\tikztotarget)}},
}

\usepackage{ifthen}
\usepackage{xparse,pgffor}

\tikzset{
mark/.code={
\tikzset{postaction={/network/mark/.cd,#1,/tikz/.cd,decorate,decoration={name=markings,mark=at position \netmarkpos with{
\begin{scope}[netmarktrafo]
\netmarkcode
\end{scope}
}}}}
\def\netmarkpos{0.5}
},
}

\def\netmarkpos{0.5}
\def\netmarkposoff{0cm}
\def\netmarkcode{}

\tikzset{
netmarktrafo/.style={},
netmarkstyle/.style={solid,semithick,sharp corners},
}

\pgfkeys{
/network/mark/.cd,
sty/.code={
\tikzset{netmarkstyle/.style={#1}}
},
asty/.code={
\tikzset{netmarkstyle/.append style={#1}}
},
f/.style={asty=fill},
p/.code={ 
\def\netmarkpos{#1}
},
poff/.code={ 
\def\netmarkposoff{#1cm}
},
e/.code={ 
\def\netmarkpos{\pgfdecoratedpathlength-0.005cm-\netmarkposoff}

\tikzset{netmarktrafo/.append style={shift={(-\netmarkwidth,0)}}}
},
s/.code={ 
\def\netmarkpos{0.005cm+\netmarkposoff}

\tikzset{netmarktrafo/.append style={shift={(\netmarkwidth,0)},xscale=-1,yscale=-1}}
},
a/.code={ 
\def\netmarkpos{\pgfdecoratedpathlength-0.005cm}

\tikzset{netmarktrafo/.append style={xscale=-1,shift={(-\netmarkwidth,0)}}}
},
b/.code={ 
\def\netmarkpos{0.005cm}

\tikzset{netmarktrafo/.append style={xscale=-1,shift={(\netmarkwidth,0),yscale=-1}}}
},
-/.code={ 
\tikzset{netmarktrafo/.append style={xscale=-1}}
},
r/.code={ 
\tikzset{netmarktrafo/.append style={yscale=-1}}
},
sideoff/.code={ 
\tikzset{netmarktrafo/.append style={shift={(0,#1)}}}
},
slab/.code={ 
\def\netmarkwidth{0}
\def\netmarkcode{
\node[inner sep=0.04cm,netmarkstyle,draw=none] (mylabelwidthtest) at (0,0){\phantom{#1}};
\path let \p1=(mylabelwidthtest.north east), \p2=(mylabelwidthtest.south east), \n1 = {max(abs(\y1),abs(\y2))} in node[inner sep=0.04cm,netmarkstyle] at (0,\n1) {#1};
}
},
lab/.code={ 
\def\netmarkwidth{0}
\def\netmarkcode{
\node[inner sep=0.04cm,anchor=\netmarkanchor] (mylabelwidthtest) at (0,0) {\phantom{#1}};
\draw[white] (mylabelwidthtest.\pgfdecoratedangle)--(mylabelwidthtest.\pgfdecoratedangle+180);
\node[inner sep=0.04cm,anchor=\netmarkanchor,netmarkstyle] at (0,0) {#1};
}
},
rotlab/.code={ 
\def\netmarkwidth{0}
\def\netmarkcode{
\node[inner sep=0.04cm,fill=white,transform shape,rotate=90,anchor=\netmarkrotanchor,netmarkstyle] (mydecorationnodename) at (0,0) {#1};
}
},
mrotlab/.style={ 
rotlab={$\scriptstyle{#1}$}
},
arr/.code={ 
\def\netmarkwidth{0.04}
\def\netmarkcode{\draw[netmarkstyle] (-0.04,0.08)--(0.04,0)--(-0.04,-0.08);}
},
arrbar/.code={ 
\def\netmarkwidth{0.08}
\def\netmarkcode{\draw[netmarkstyle] (-0.08,0.08)--(0,0)--(-0.08,-0.08) (0.04,0.08)--(0.04,-0.08);}
},
rarr/.code={ 
\def\netmarkwidth{0.04}
\def\netmarkcode{\draw[netmarkstyle] (-0.04,-0.08)arc(90-180:90:0.08);}
},
dot/.code={ 
\def\netmarkwidth{0.08}
\def\netmarkcode{\draw[netmarkstyle] (0,0)circle(0.08);}
},
flag/.code={ 
\def\netmarkwidth{0.06}
\def\netmarkcode{\draw[netmarkstyle] (-0.06,0)--(0,0.09)--(0.06,0)--cycle;}
},
darr/.code={ 
\def\netmarkwidth{0.08}
\def\netmarkcode{\draw[netmarkstyle] (-0.04,0)--(0.04,0)--(-0.04,0.08)--cycle;}
},
circ/.code={
\def\netmarkwidth{0.1}
\def\netmarkcode{\draw[netmarkstyle] (0,0) circle (0.1);}
},
hcirc/.code={ 
\def\netmarkwidth{0.1}
\def\netmarkcode{\draw[netmarkstyle] (-0.1,0) arc (180:0:0.1);}
},
diamond/.code={
\def\netmarkwidth{0.1}
\def\netmarkcode{\draw[netmarkstyle] (-0.1,0)--(0,-0.1)--(0.1,0)--(0,0.1)--cycle;}
},
bar/.code={ 
\def\netmarkwidth{0.05}
\def\netmarkcode{
\draw[netmarkstyle] (0,-0.08cm-0.5*\pgflinewidth)--(0,0.08cm+0.5*\pgflinewidth);
}
},
dbar/.code={ 
\def\netmarkwidth{0.13}
\def\netmarkcode{
\draw[netmarkstyle] (-0.04cm,-0.08cm-0.5*\pgflinewidth)--(-0.04cm,0.08cm+0.5*\pgflinewidth) (0.04cm,-0.08cm-0.5*\pgflinewidth)--(0.04cm,0.08cm+0.5*\pgflinewidth);
}
},
hbar/.code={ 
\def\netmarkwidth{0.05}
\def\netmarkcode{
\draw[netmarkstyle] (0, 0.5*\pgflinewidth)--++(0,0.12);
}
},
mill/.code={
\def\netmarkwidth{0.16}
\def\netmarkcode{
\draw[netmarkstyle] (0,-0.5*\pgflinewidth)--++(-0.08,-0.08)--++(0,0.08);
\draw[netmarkstyle] (0,0.5*\pgflinewidth)--++(0.08,0.08)--++(0,-0.08);
}
},
three dots/.code={
\def\netmarkwidth{0.2}
\def\netmarkcode{
\fill (-0.12,0) circle (0.5*0.05) (0,0) circle (0.5*0.05) (0.12,0) circle (0.5*0.05);
}
},
}

\tikzset{wid/.style={minimum width=#1cm}}
\tikzset{hei/.style={minimum height=#1cm}}

\tikzset{sx/.style={xshift=#1cm}}
\tikzset{sy/.style={yshift=#1cm}}

\tikzset{box/.style={draw,rectangle}}
\tikzset{fbox/.style={draw,rectangle, line width=1.1}}
\tikzset{roundbox/.style={draw,rectangle,rounded corners}}
\tikzset{froundbox/.style={draw,rectangle, rounded corners, line width=1.1}}
\tikzset{rounddiamond/.style={draw,diamond,rounded corners}}
\tikzset{dot/.style={draw, shape=circle, fill=black, scale=0.5}}

\tikzset{
netbox/.code={
\node[draw,netbdstyle] (\atomname) at (0,0) {#1};
\coordinate (\atomname-r) at (\atomname.east);
\coordinate (\atomname-l) at (\atomname.west);
\coordinate (\atomname-t) at (\atomname.north);
\coordinate (\atomname-b) at (\atomname.south);
\coordinate (\atomname-tr) at (\atomname.north east);
\coordinate (\atomname-br) at (\atomname.south east);
\coordinate (\atomname-tl) at (\atomname.north west);
\coordinate (\atomname-bl) at (\atomname.south west);
},
}

\tikzset{bdlw/.code={\tikzset{mybdstyle/.style={draw, line width=#1}}}}
\tikzset{bdcol/.code={\tikzset{mybdstyle/.append style={#1}}}}

\newcommand\atoms[2]{
\foreach \name/\keys in {#2}{
\expandafter\atom\expandafter{\keys,#1}{\name}
}
}

\newcommand\atom[2]{
\def\atomname{#2}
\tikzset{
nettrafo/.style={},
netatompos/.style={},
netdeco/.style={},
netpostdeco/.style={},
}

\pgfkeys{/network/atom/.cd,#1}

\begin{scope}[netatompos] 
\begin{scope}[nettrafo] 
\netshapecoords 
\fill[netbackstyle] \netshapepath;
\clip \netshapepath;
\tikzset{netdeco}
\draw[netbdstyle] \netshapepath;
\end{scope}
\tikzset{netpostdeco} 
\end{scope}

}

\pgfkeys{
/network/atom/.cd,
square/.code={

\def\netshapepath{(-\tempsize,-\tempsize)rectangle (\tempsize,\tempsize)}
\def\netshapecoords{
\node[rectangle,wid=2*\tempsize,hei=2*\tempsize,inner sep=0,transform shape](\atomname)at(0,0){};
\coordinate(\atomname-c) at (0,0);
\coordinate(\atomname-r) at (\tempsize,0);
\coordinate(\atomname-l) at (-\tempsize,0);
\coordinate(\atomname-t) at (0,\tempsize);
\coordinate(\atomname-b) at (0,-\tempsize);
\coordinate(\atomname-br) at (\tempsize,-\tempsize);
\coordinate(\atomname-tr) at (\tempsize,\tempsize);
\coordinate(\atomname-bl) at (-\tempsize,-\tempsize);
\coordinate(\atomname-tl) at (-\tempsize,\tempsize);
}},
circ/.code={

\def\netshapepath{(0,0)circle(\tempsize)}
\def\netshapecoords{
\node[circle,wid=2*\tempsize,hei=2*\tempsize,inner sep=0,transform shape](\atomname)at(0,0){};
\coordinate(\atomname-c) at (0,0);
\coordinate(\atomname-r) at (\tempsize,0);
\coordinate(\atomname-l) at (-\tempsize,0);
\coordinate(\atomname-t) at (0,\tempsize);
\coordinate(\atomname-b) at (0,-\tempsize);
}},
triang/.code={

\def\netshapepath{(-30:\tempsize)--(90:\tempsize)--(-150:\tempsize)--cycle}
\def\netshapecoords{
\node[regular polygon,regular polygon sides=3,wid=2*\tempsize,inner sep=0,transform shape](\atomname)at(0,0){};
\coordinate(\atomname-c) at (0,0);
\coordinate(\atomname-cr) at (-30:\tempsize);
\coordinate(\atomname-cl) at (-150:\tempsize);
\coordinate(\atomname-ct) at (90:\tempsize);
\coordinate(\atomname-mb) at (-90:0.5*\tempsize);
\coordinate(\atomname-mr) at (30:0.5*\tempsize);
\coordinate(\atomname-ml) at (150:0.5*\tempsize);
}},
diamond/.code={

\def\netshapepath{(0,-\tempsize)--(\tempsize,0)--(0,\tempsize)--(-\tempsize,0)--cycle}
\def\netshapecoords{
\node[rotate=45,rectangle,wid=sqrt(2)*\tempsize,hei=sqrt(2)*\tempsize,inner sep=0,transform shape](\atomname)at(0,0){};
\coordinate(\atomname-c) at (0,0);
\coordinate(\atomname-r) at (\tempsize,0);
\coordinate(\atomname-l) at (-\tempsize,0);
\coordinate(\atomname-t) at (0,\tempsize);
\coordinate(\atomname-b) at (0,-\tempsize);
}},
pentagon/.code={

\def\netshapepath{(-126:\tempsize)--(-54:\tempsize)--(18:\tempsize)--(90:\tempsize)--(162:\tempsize)--cycle}
\def\netshapecoords{
\node[regular polygon,regular polygon sides=5,wid=2*\tempsize,inner sep=0,transform shape](\atomname)at(0,0){};
\coordinate(\atomname-c) at (0,0);
\coordinate (\atomname-mb)at(-90:{\tempsize*cos(36)});
\coordinate (\atomname-mbr)at(-18:{\tempsize*cos(36)});
\coordinate (\atomname-mtr)at(54:{\tempsize*cos(36)});
\coordinate (\atomname-mtl)at(126:{\tempsize*cos(36)});
\coordinate (\atomname-mbl)at(-162:{\tempsize*cos(36)});
\coordinate (\atomname-cbr)at(-54:\tempsize);
\coordinate (\atomname-cr)at(18:\tempsize);
\coordinate (\atomname-ct)at(90:\tempsize);
\coordinate (\atomname-cl)at(162:\tempsize);
\coordinate (\atomname-cbl)at(-126:\tempsize);
}},
halfcirc/.code={

\def\netshapepath{(\tempsize,0)arc(0:180:\tempsize)--++(0,-0.04)-|cycle}
\def\netshapecoords{
\node[circle,wid=2*\tempsize,hei=2*\tempsize,inner sep=0,transform shape](\atomname)at(0,0){};
\coordinate(\atomname-c) at (0,0);
\coordinate(\atomname-r) at (\tempsize,0);
\coordinate(\atomname-l) at (-\tempsize,0);
\coordinate(\atomname-t) at (0,\tempsize);
\coordinate(\atomname-b) at (0,0);
}},
void/.code={
\def\netshapepath{}
\def\netshapecoords{
\coordinate(\atomname) at (0,0);
\coordinate(\atomname-c) at (0,0);
}},
labbox/.code={
\def\netshapepath{(0,0)}
\def\netshapecoords{}
\tikzset{netpostdeco/.append style={netbox=#1}}
},
}

\pgfkeys{
/network/atom/.cd,
p/.code={\tikzset{netatompos/.append style={shift={(#1)}}}},
xscale/.code={\tikzset{nettrafo/.append style={xscale=#1}}},
yscale/.code={\tikzset{nettrafo/.append style={yscale=#1}}},
scale/.code={\tikzset{nettrafo/.append style={scale=#1}}},
rot/.code={\tikzset{nettrafo/.append style={rotate=#1}}},
hflip/.style={xscale=-1},
vflip/.style={yscale=-1},
big/.style={scale=3/2},
small/.style={scale=2/3},
huge/.style={scale=9/4},
tiny/.style={scale=4/9},
flat/.style={yscale=2/3},
fflat/.style={yscale=4/9},
}

\tikzset{
netbdstyle/.style={line width=0.15em}, 
netdecstyle/.style={},
netpostdecstyle/.style={},
netbackstyle/.style={white},
}
\pgfkeys{
/network/atom/.cd,
bdstyle/.code={\tikzset{netbdstyle/.style={#1}}},
bdastyle/.code={\tikzset{netbdstyle/.append style={#1}}},
decstyle/.code={\tikzset{netdecstyle/.style={#1}}},
decastyle/.code={\tikzset{netdecstyle/.append style={#1}}},
postdecstyle/.code={\tikzset{netpostdecstyle/.style={#1}}},
postdecastyle/.code={\tikzset{netpostdecstyle/.append style={#1}}},
backstyle/.code={\tikzset{netbackstyle/.style={#1}}},
backastyle/.code={\tikzset{netbackstyle/.append style={#1}}},
style/.style={bdstyle=#1, decstyle=#1, postdecstyle=#1},
astyle/.style={bdastyle=#1, decastyle=#1, postdecastyle=#1},
whiteblack/.style={decstyle=white,backstyle=black},
nobd/.style={bdstyle={line width=0}},
}

\tikzset{
netbscope/.code={\begin{scope}[#1]},
netescope/.code={\end{scope}},
}

\pgfkeys{
/network/atom/.cd,
bscope/.code={\tikzset{netdeco/.append style={netbscope={#1}}}},
escope/.code={\tikzset{netdeco/.append style={netescope}}},
dec/.style 2 args={bscope={#1},#2,escope}, 
vbar/.style={dec={rotate=90}{hbar}},
dcross/.style={dec={rotate=45}{hbar},dec={rotate=-45}{hbar}},
cross/.style={hbar,vbar},
sect6/.style={sect=60},
sect8/.style={sect=45},
sect10/.style={sect=36},
}

\def\regdec#1{\pgfkeys{/network/atom/.cd,#1/.code={\tikzset{netdeco/.append style={net#1}}}}}
\regdec{all}\regdec{rhalf}\regdec{rquart}\regdec{brquart}\regdec{dot}\regdec{spiral}\regdec{swirl}\regdec{hstripe}\regdec{hbar}\regdec{rrey}
\pgfkeys{/network/atom/.cd,sect/.code={\tikzset{netdeco/.append style={netsect=#1}}}}

\tikzset{
netall/.code={\fill[netdecstyle] (-0.3,-0.3)rectangle (0.3,0.3);}, 
netrhalf/.code={\fill[netdecstyle] (0,-0.3)rectangle (0.3,0.3);}, 
netrquart/.code={\fill[netdecstyle] (0.075,-0.3)rectangle (0.3,0.3);}, 
netbrquart/.code={\fill[netdecstyle] (0,0)rectangle (0.3,-0.3);}, 
netsect/.code={\fill[netdecstyle] (0,0)--(0,-0.3)arc(-90:-90+#1:0.3)--cycle;}, 
netdot/.code={\fill[netdecstyle] (0,0)circle(0.07);}, 
netspiral/.code={\draw[netdecstyle] plot [variable=\t,domain=0:4] ({0.075*\t*cos(pi*(\t-0.5) r)},{0.075*\t*sin(pi*(\t-0.5) r)});}, 
netswirl/.code={\fill[netdecstyle] plot [variable=\t,domain=0:2] ({0.15*\t*cos(pi*(\t-0.5) r)},{0.15*\t*sin(pi*(\t-0.5) r)}) arc(-90:-450:0.3)--cycle;}, 
nethstripe/.code={\fill[netdecstyle] (-0.3,-0.05)rectangle(0.3,0.05);}, 
nethbar/.code={\draw[netdecstyle] (-0.3,0)--(0.3,0);}, 
netrrey/.code={\draw[netdecstyle] (0,0)--(0.3,0);} 
}

\pgfkeys{
/network/atom/.cd,
postbscope/.code={\tikzset{netpostdeco/.append style={netbscope={#1}}}},
postescope/.code={\tikzset{netpostdeco/.append style={netescope}}},
postdec/.style 2 args={postbscope={#1},#2,postescope}, 
arc/.code={\tikzset{netpostdeco/.append style={netarc=#1}}},
lab/.code={\tikzset{netpostdeco/.append style={netlab={#1}}}},
shadecirc/.code={\tikzset{netpostdeco/.append style=netshadecirc}},
shaderect/.code={\tikzset{netpostdeco/.append style={netshaderect={#1}}}},
debug/.code={\tikzset{netpostdeco/.append style=netdebug}},
markline/.code 2 args={\tikzset{netpostdeco/.append style={netmarkline={#1}{#2}}}},
postcirc/.code={\tikzset{netpostdeco/.append style=netpostcirc}},
}

\tikzset{
netlab/.code={
\pgfkeys{/network/atom/lab/.cd,#1}
\node[netpostdecstyle] at (\ifdefined\netlabpos\netlabpos\else\netlabang:\netlabdist\fi) {\netlabwrap{\netlabtext}};
},
netarc/.code args={#1:#2:#3}{
\draw[netpostdecstyle] (#1:#3) arc (#1:#2:#3);
},
netshadecirc/.code= {
\fill[opacity=0.4,netpostdecstyle] (0,0)circle(0.4);
},
netpostcirc/.code= {
\draw[netpostdecstyle] (0,0)circle(0.15);
},
netshaderect/.code= {
\fill[rc,opacity=0.4,netpostdecstyle] ($-1*(#1)$) rectangle (#1);
},
netdebug/.code= {
\node[red] at (0,0){\atomname};
},
netmarkline/.code 2 args= {
\draw (\atomname)edge[mark={#2}]++(#1);
},
}

\def\netlabwrap#1{#1}
\pgfkeys{
/network/atom/lab/.cd,
t/.code={\def\netlabtext{#1}}, 
p/.code={\def\netlabpos{#1}}, 
ang/.code={\def\netlabang{#1}},
dist/.code={\def\netlabdist{#1}},
wrap/.code={\def\netlabwrap##1{#1}}, 
}

\usepackage{needspace}
\usepackage{braket}
\usepackage{bbold,bm}
\usepackage{amssymb,amsmath,mathtools}
\usepackage[ruled]{algorithm}
\usepackage{algpseudocode}
\usepackage{diagbox}
\usepackage{hyperref}

\newcommand\myparagraph[1]{\noindent\textbf{#1 ---}}

\newcommand\coeffs[1]{\begin{vmatrix}#1\end{vmatrix}}
\newcommand{\mpm}[1]{\begin{pmatrix}#1\end{pmatrix}}

\def\zz{\mathbb Z}
\def\rr{\mathbb R}
\def\ovl{\overline}
\def\img{\operatorname{img}}
\def\idop{\operatorname{id}}
\def\gphys{G_{\text{phys}}}
\def\mmod{\operatorname{mod}}
\def\pullb{\bm *}
\def\checkx{A}
\def\nrxch{m}
\def\checkz{B}
\def\xdecode{C}
\def\nrzch{o}

\tikzset{
ind/.style={mark={lab=#1,a}}, 
startind/.style={mark={lab=#1,b}}, 
}

\begin{document}
\title{Finding diagonal logical gates in CSS codes and circuits}
\author{Andreas Bauer}
\email{andib@mit.edu}
\affiliation{\footnotesize Department of Mechanical Engineering, Massachusetts Institute of Technology, Cambridge, MA 02139, USA}

\begin{abstract}
Finding efficient schemes for non-Clifford logic or magic state preparation is one of the central challenges on the way to fault-tolerant quantum computation.
Many of the proposed schemes rely on diagonal non-Clifford logical gates acting on CSS codes in space or decorating CSS-type syndrome-extraction circuits in spacetime.
Here we propose and implement efficient algorithms to find all (spacetime) logical gates of a given CSS code (circuit) composed from a prescribed set of ansatz gates.
Depending on the choice of ansatz gates, this means finding transversal gates, more general locality-preserving logical circuits, folding gates, or similar.
While we focus on qubit diagonal gates in the Clifford hierarchy, we also discuss the generalization to arbitrary diagonal non-hierarchy gates, certain non-diagonal gates, as well as prime and composite-dimensional qudits.
Our method works by rephrasing code-space preserving gates as the kernel of the ``pullback'' of the $X$ check matrix onto phase functions, which maps between finite abelian 2-groups.
We implement a fast ``filtration'' method to find this kernel.
The runtime for finding fault-tolerant logical gates in a qLDPC code with $O(n)$ qubits or a circuit with $O(n)$ gates in a naive dense implementation is $O(n^3)$, with potential for improvement making use of sparsity.
\end{abstract}

\maketitle
\tableofcontents

\section{Introduction}
\myparagraph{Motivation}
Large-scale universal quantum computers need to fault-tolerantly implement both Clifford and non-Clifford operations.
While Clifford operations are often easy to implement in the usual Pauli-stabilizer based language of quantum error correction, non-Clifford gates pose more of a challenge.
There are at least four approaches to this challenge, all of which relate to non-Clifford transversal logical gates.
One approach is magic state distillation~\cite{Bravyi2004,Bravyi2012,Litinski2019,Wills2024}, which produces a higher-fidelity magic state from many lower-fidelity copies.
Distillation protocols are often derived by executing a non-Clifford transversal gate on a \emph{distillation code} via gate teleportation using the lower-fidelity magic states.
A second approach is to use transversal non-Clifford gates directly.
Even though transversal gates alone cannot be universal~\cite{Eastin2008}, this can be achieved using a ``dimension jump'' or more general code switching~\cite{Bombin2014,Paetznick2013,Kubica2014,Beverland2021,Zhu2023,Li2025}.
A third approach is to directly measure a non-local transversal Clifford (non-Pauli) operator, a method known as \emph{magic state cultivation}~\cite{Chamberland2020,Bombin2022,Gidney2024,Rosenfeld2025}.
Though not scalable, this method can produce magic states at a fidelity that may be high enough for certain quantum algorithms in practice, or may serve as a starting point for further distillation.
For measuring a Clifford gate we need a circuit containing non-Clifford gates, and these gates can be viewed as ``spacetime logical'' in a sense that will be explained in Section~\ref{sec:spacetime_logical}.
A fourth approach is to insert ``spacetime logical'' non-Clifford gates into a backbone Clifford circuit for an extended amount of time.
The first such protocols were developed by taking 3+0-dimensional dimension-jump protocols and turning one space direction into a time direction~\cite{Bombin2018,Brown2019}.
Later it was realized that such protocols implement non-Abelian topological phases~\cite{Davydova2025,twisted_double_code,twisted_color_circuits,non_clifford_benchmarking}.
However, the method also applies to more general qLDPC codes, and examples have been constructed, for example from a perspective of gauging~\cite{Zhu2026,Williamson2026,Christos2026}.
These protocols can also be viewed as a scalable version of magic state cultivation.

Due to their significance, a lot of effort has been put recently towards finding codes with transversal non-Clifford gates.
The predominant class of examples that has been considered are CSS/qLDPC codes with diagonal transversal non-Clifford gates such as $T$ or $CCZ$.
Notable examples include Refs.~\cite{Zhu2023,Breuckmann2024,Lin2024,Li2026}.

\myparagraph{Contributions}
In this work, we propose and implement an efficient algorithm to find all locality-preserving constant-depth diagonal gates of a given CSS code, including non-Clifford ones.
So this includes transversal gates, but also gates that do not act on-site or on non-overlapping sets of qubits.
We also do not impose any notion of translation invariance - and therefore include gates corresponding to ``higher-order symmetries''.

More precisely, we find all logical gates composed from a prescribed set of diagonal \emph{ansatz gates}.
Each ansatz gate is specified by (1) a subset of $i$ qubits for some (small) number $i$ and (2) a subgroup of the group $\simeq U(1)^{2^i}$ of all diagonal gates acting on the $i$ qubits.
The group of all considered physical gates is thus the product of all individual subgroups.
Our algorithm finds the subgroup of the considered gates that preserve the code space.

If we choose the ansatz gates such that each qubit participates only in $l$ of them, then the considered gates are circuits of ansatz gates of depth at most $l$.
For a qLDPC code family with degree-bounded Tanner graphs, a natural choice is to include all local gates as ansatz gates.
That is, we choose some small constant $r$, and include each ansatz gate inside a ball of radius $<r$.
That way we can find any code-space preserving gate with the same locality structure as the CSS code itself.
We can also find non-local gates by using ansatz gates that couple qubits that are not nearby in the Tanner graph.
This includes gates that are only local after ``stacking'' or ``folding''.
Note that in this case we need to prescribe how the code(s) is/are folded/stacked - the algorithm does not consider all possible ways of stacking/folding.

The starting point for our method is the fact that diagonal logical gates that preserve the code space of a CSS code must assign the same phase to every $Z$-basis configuration supporting the code space.
In particular, the phase that a logical diagonal gate assigns to a qubit configuration must be trivial for configurations in the image of the $X$-check matrix $A$.
We can consider the pullback $A^{\pullb}$ of the $X$-check matrix $A$, which maps phase functions on the qubit configurations to phase functions on the $X$-check configurations.
Logical gates are the ones that yield a trivial phase function on the $X$ checks, that is, they are in the kernel of $A^{\pullb}$.
Thus finding all logical gates can be reduced to finding the kernel of the abelian group homomorphism $A^{\pullb}$, which can be done efficiently.
In addition to the efficient algorithm, our formalism also provides a clean understanding of the structure of diagonal transversal or locality-preserving logical gates in CSS codes, and can be seen as a generalization of the triorthogonality condition~\cite{Bravyi2012}.

Our algorithm can also be used to find \emph{spacetime local logical gates} as we discuss in Section~\ref{sec:spacetime_logical}.
A spacetime local logical gate is a way to decorate a CSS-type syndrome-extraction circuit with diagonal (non-Clifford) gates in a fault-tolerant way.
This includes the scalable protocols of Refs.~\cite{Davydova2025,twisted_color_circuits,twisted_double_code}, and also magic state cultivation~\cite{Chamberland2020,Bombin2022,Gidney2024,Rosenfeld2025}.
It also includes ways to interleave the application of an ordinary transversal/locality-preserving gate with syndrome extraction.
Our method also applies to certain non-diagonal gates, as they can be interpreted as phase terms coupling qubit configurations of one time step in the circuit with the next.
Independent of that, we can also find gates that are diagonal in the $X$ basis.

The fact that we can freely choose the ansatz gates makes our method particularly useful on the circuit level:
We can put ansatz gates at locations in the circuit that are particularly convenient from a hardware perspective.
So the approach can not only be used to find new logical gates, but also to find good circuit implementations for known logical gates.

While on-site transversal non-Clifford gates only exist for rather special CSS codes, having constant-depth locality-preserving logical gates is far more generic.
In particular, if two qLDPC codes are related by locality-preserving cohomological chain maps, then their locality-preserving logical gates are in one-to-one correspondence.
The most famous example for this is perhaps the equivalence between the 3D color code and three copies of the 3D toric code:
The transversal $T$ gate of the color code corresponds to a transversal $CCZ$ gate of the toric code~\cite{Kubica2015}.
It is thus conceivable that the method can be used to discover new qLDPC codes that possess a locality-preserving logical gate.

Our method is particularly useful for extending a logical gate that exists in one region of a CSS code or circuit to other regions.
For example, consider a ``bulk'' qLDPC code with a logical gate, and some ``boundary'' for the code.
Then we can efficiently find ways to extend that logical gate to the boundary, which might not always be obvious.
Similarly, consider two qLDPC codes with logical gates that are interfaced along some ``domain wall''.
Then we can find ways to ``blend'' the two different logical gates into each other, that is, extend the two different logical gates  to the common domain wall.

Finally, we provide a fast python implementation of the proposed algorithm, for the case where the ansatz gates are in the Clifford hierarchy~\cite{diagonal_gate_finder}.
We develop a subroutine called ``filtration'' to efficiently find the kernel of the group homomorphism $A^{\pullb}$, which maps between two abelian 2-groups, that is, groups of the form $\zz_2^\bullet\times \zz_4^\bullet\times \zz_8^\bullet\times \ldots$.
The subroutine reduces the problem to finding multiple kernels of $\zz_2$-valued matrices, which can be implemented by Gaussian elimination in a fast bit-packed manner.

\myparagraph{Literature review}
Let us put our method into context relative to the existing literature.
One of the first mathematical conditions on codes used to construct transversal gates is the \emph{triorthogonality condition} introduced in Ref.~\cite{Bravyi2012}.
This condition guarantees that the code has a transversal gate that acts like $T$ on every qubit, in addition to some $S$ or $CZ$ terms.
The condition is later generalized in Ref.~\cite{Haah2017}.
Our formalism can be viewed as a further generalization of the triorthogonality condition to gates that are arbitrary products of Clifford-hierarchy gates.

There have also been several references that propose numerical algorithms for finding transversal gates.
Ref.~\cite{Webster2023} gives a computational method for efficiently finding transversal non-overlapping product diagonal Clifford-hierarchy gates of CSS codes, but they do not consider more general constant-depth locality-preserving gates.
Their method is based on the XP formalism and is quite different from ours.
Refs.~\cite{Rengaswamy2019,Sayginel2024} give algorithms for calculating transversal Clifford gates on CSS and general stabilizer codes, but do not study non-Clifford operations.
Ref.~\cite{Benhemou2026} efficiently finds transversal $CX$ gates coupling two codes.
The mindset is somewhat similar to ours in that logical gates form an abelian subgroup of prescribed ansatz gates, but they do not consider non-Clifford gates.
In fact, our method can find the same $CX$ gates as they become $CZ$ gates after a $Z\leftrightarrow X$ duality transformation on the target code.

\myparagraph{Structure}
In Section~\ref{sec:theory}, we formulate the general theory and algorithm, focussing on gates in the third order of the Clifford hierarchy.
In Section~\ref{sec:generalizations} we introduce various generalizations.
Most importantly, we discuss the generalization to spacetime local logical gates in Section~\ref{sec:spacetime_logical}.
We also discuss higher Clifford levels, arbitrary diagonal gates, qudit CSS codes of prime or composite dimensions, and certain non-diagonal gates in Sections~\ref{sec:higher_clifford}, \ref{sec:arbitrary_diagonal}, \ref{sec:qudits}, and \ref{sec:nondiagonal_nonclifford}.
We also discuss how making use of translation invariance or $k$-locality of qLDPC codes could be used to further speed up the algorithm in Section~\ref{sec:faster_locality}.

\section{Theory and algorithm}
\label{sec:theory}
In this section, we describe the algorithm to efficiently find all diagonal logical gates of a given CSS code consisting of a subset of prescribed ansatz gates.
For concreteness and simplicity, we focus on gates in the third level of the Clifford hierarchy~\cite{Gottesman1999, Cui2016}, that is, products of $T$, $CS$, and $CCZ$ gates.
In other words, the ansatz gates are single qubits with a $\zz_8$ subgroup of diagonal gates generated by $T$, qubit pairs with a $\zz_4$ subgroup generated by $CS$, and qubit triples with a $\zz_2$ subgroup generated by $CCZ$ gates.
Arbitrary Clifford hierarchy gates and various other generalizations are described in Section~\ref{sec:generalizations}.

\subsection{Diagonal logical gates}
\myparagraph{CSS codes}
A \emph{CSS code} on $n$ qubits is defined by a binary $n\times \nrxch$ matrix $\checkx$ (the $X$-check matrix) and an $\nrzch\times n$ matrix $\checkz$ (the transpose of the $Z$-check matrix) fulfilling $\checkz \checkx = 0$ (such that $X$ and $Z$ checks commute).
$\checkx$ and $\checkz$ are the coefficient matrices of group homomorphisms that we denote by the same symbols,
\begin{equation}
\checkx: \zz_2^{\nrxch}\rightarrow \zz_2^n\;,\qquad \checkz: \zz_2^n\rightarrow \zz_2^\nrzch\;,
\end{equation}
and form a chain complex
\begin{equation}
\zz_2^{\nrxch}\xrightarrow{\checkx} \zz_2^n\xrightarrow{\checkz} \zz_2^\nrzch\;.
\end{equation}
CSS codes have a canonical logical basis labeled by the cohomology classes of the chain complex, that is, elements of the quotient $[a]\in \ker(\checkz) / \img(\checkx)$.
The code state for the cohomology class $[a]$ is an equal-weight superposition of all cycles in $[a]$:
\begin{equation}
\label{eq:code_state}
\ket{[a]} = \sum_{\substack{a: \checkz a=0,\\a\in [a]}} \ket a
= \sum_{x\in\zz_2^{\nrxch}} \ket{a_0+\checkx x}
\;,
\end{equation}
where $a_0\in \zz_2^n$ is any fixed cycle in $[a]$, that is, any element such that $\checkz a_0=0$ and $a_0\in [a]$.

\myparagraph{Third-order diagonal gates}
We consider gates $V_{\mathbf c}$ specified by a set of coefficients
\begin{equation}
\begin{gathered}
\mathbf c\coloneqq (\{c^{(0)}_{ijk}\}_{(i,j,k)\in N^{(0)}},\{c^{(1)}_{ij}\}_{(i,j)\in N^{(1)}},\{c^{(2)}_i\}_{i\in N^{(2)}})\\
\in \gphys \coloneqq \zz_2^{N^{(0)}}\times \zz_4^{N^{(1)}} \times \zz_8^{N^{(2)}}\;,
\end{gathered}
\end{equation}
where $N^{(0)}$ is a subset of qubit triples, $N^{(1)}$ is a subset of qubit pairs, and $N^{(2)}\subset\{0,\ldots,n-1\}$ is a subset of individual qubits.
The triples and pairs in $N^{(0)}$ and $N^{(1)}$ do not need to be disjoint.
Explicitly, $V_{\mathbf c}$ is given by
\begin{equation}
V_{\mathbf c}
\coloneqq \prod_{(i,j,k)\in N^{(0)}} CCZ_{ijk}^{c^{(0)}_{ijk}} \prod_{(i,j)\in N^{(1)}} CS_{ij}^{c^{(1)}_{ij}} \prod_{i\in N^{(2)}} T_i^{c^{(2)}_i}\;.
\end{equation}
In other words, $V_{\mathbf c}$ is a product of $CCZ^c$ on the qubit triples in $N^{(0)}$, $CS^c$ gates on the qubit pairs in $N^{(1)}$, and $T^c$ gates on the qubits in $N^{(2)}$.
Note that we could set $N^{(0)}$, $N^{(1)}$, and $N^{(2)}$ to be the full subsets of all qubit triples, qubit pairs, and individual qubits and efficiently find all third-order diagonal code-space preserving gates.
However, these gates would include non-locality-preserving/non-fault-tolerant gates where a large number of different gates act on a single qubit.
By choosing $N^{(0)}$ and $N^{(1)}$ such that each qubit is only contained in a constant (small) number of triples/pairs, we can restrict ourselves to gates that are constant-depth and can be implemented in a scalable fault-tolerant manner.

Consider any integer $n$ and collection of triples, pairs, and individuals $\mathbf N\coloneqq (N^{(0)}, N^{(1)}, N^{(2)})$ of the set $\{0,\ldots,n-1\}$.
For any set of coefficients $\mathbf c$ as above, define the function
\begin{equation}
\label{eq:third_order_function}
\begin{gathered}
S^{n,\mathbf N}_{\mathbf c}: \zz_2^n\rightarrow \rr/\zz\;,\\
\begin{multlined}
S^{n,\mathbf N}_{\mathbf c}(a)
\coloneqq \frac12 \sum_{(i,j,k)\in N^{(0)}} c^{(0)}_{ijk} \ovl a_i \ovl a_j \ovl a_k\\
+ \frac14 \sum_{(i,j)\in N^{(1)}} c^{(1)}_{ij} \ovl a_i \ovl a_j
+ \frac18 \sum_{i\in N^{(2)}} c^{(2)}_i \ovl a_i\;,
\end{multlined}
\end{gathered}
\end{equation}
which we call the \emph{third-order function} for $\mathbf c$.
Here, $\rr/\zz$ denotes the group of real numbers modulo $1$, represented by the interval $[0,1)$, which is isomorphic to the circle group $U(1)$ of complex phases (but written additively).
For $a\in\zz_2$, $\ovl a\in \zz$ denotes the canonical embedding (or ``lift'') of $\zz_2\simeq \{0,1\}$ into $\zz$, and it will become important to distinguish between $a$ and $\ovl a$ in the following sections.
We provide a natural algebraic explanation of why the above functions are ``third-order'' in Section~\ref{sec:qudits}.
Note that we will sometimes drop the superscripts $n,\mathbf N$ when clear from the context.
Using the function $S_{\mathbf c}$, the action of $V_{\mathbf c}$ on a qubit computational basis vector $\ket a$ can be written as
\begin{equation}
V_{\mathbf c}\ket a= e^{2\pi i S_{\mathbf c}(a)} \ket a\;,
\end{equation}
Multiplying gates and adding third-order functions $S$ is compatible with adding the coefficient collections $\mathbf c\in \gphys$,
\begin{equation}
\label{eq:s_linear}
V_{\mathbf c_0} V_{\mathbf c_1}=V_{\mathbf c_0+\mathbf c_1}\;,\qquad S_{\mathbf c_0}+S_{\mathbf c_1}=S_{\mathbf c_0+\mathbf c_1}\;.
\end{equation}

\myparagraph{Code space preservation}
By definition, a logical gate preserves the code space.
Eq.~\eqref{eq:code_state} shows that a necessary and sufficient condition for this is that $V_{\mathbf c}$ contributes the same phase to each basis configuration in the same cohomology class,
\begin{equation}
\label{eq:cohomology_invariant}
\begin{gathered}
S_{\mathbf c}(a) = S_{\mathbf c}(a') \quad\forall [a],\quad a,a'\in [a]\;,\text{or}\\
S_{\mathbf c}(a+\checkx e_j) = S_{\mathbf c}(a) \quad\forall 0\leq j<\nrxch,\quad\forall a\in\ker(B)\;,
\end{gathered}
\end{equation}
where $e_j$ denotes the $j$th generator of $\zz_2^{\nrxch}$, such that $Ae_j$ is the $j$th generating $X$ check.
Now, consider the fault-tolerant setting where the CSS code is part of a scalable qLDPC family with growing distance, and the (hyper-)graph formed by the $X$ stabilizers and ansatz gates is degree bounded.
For any $j$, let $L_j$ denote the union of the supports of all ansatz gates in $N^{(0)}$, $N^{(1)}$, and $N^{(2)}$ that share at least one qubit with the support of $Ae_j$.
The size of $L_j$ is bounded by a system-size independent constant due to the degree-boundedness assumption.
Thus, for any $a\in\ker(\checkz)$, there is $\alpha$ such that $a|_{L_j}=(\checkx\alpha)|_{L_j}$, that is, $a$ is equal to a boundary $A\alpha$ restricted to the qubit subset $L_j$.
Otherwise, a qubit-wise $Z$ measurement on $L_j$ would reveal logical information, which contradicts the growing-distance assumption as $L_j$ has constant bounded size.
With this, we find
\begin{equation}
\begin{gathered}
S_{\mathbf c}(a+Ae_j)-S_{\mathbf c}(a)\\
= \sum_{\substack{\mathbf i\in N^{(0)}\cup N^{(1)}\cup N^{(2)}:\\\mathbf i\cap Ae_j\neq 0}}S^{\mathbf i}_{\mathbf c}(a+Ae_j)-S^{\mathbf i}_{\mathbf c}(a)\\
= \sum_{\substack{\mathbf i\in N^{(0)}\cup N^{(1)}\cup N^{(2)}:\\\mathbf i\cap Ae_j\neq 0}}S^{\mathbf i}_{\mathbf c}(A\alpha+Ae_j)-S^{\mathbf i}_{\mathbf c}(A\alpha)\\
= S_{\mathbf c}(A(\alpha+e_j))-S_{\mathbf c}(A\alpha)\;,
\end{gathered}
\end{equation}
where $S_{\mathbf c}^{\mathbf i}$ for $\mathbf i\in N^{(l)}$ denotes the summand of Eq.~\eqref{eq:third_order_function} with coefficient $\mathbf c_{\mathbf i}^{(l)}$.
So in the fault-tolerant setting, Eq.~\eqref{eq:cohomology_invariant} reduces to the simpler condition
\begin{equation}
(S_{\mathbf c}\circ \checkx)(\alpha) = S_{\mathbf c}(\checkx\alpha)=0\quad\forall \alpha\in\zz_2^{\nrxch}
\;.
\end{equation}
As we will show in Section~\ref{sec:constructing_coefficients}, $S_{\mathbf c}\circ \checkx$ is again a third-order function, and we have
\begin{equation}
\label{eq:pullback_def}
S^{n,\mathbf N}_{\mathbf c}\circ \checkx = S^{\nrxch,\mathbf M}_{\checkx^{\pullb} \mathbf c}
\end{equation}
for some function
\begin{equation}
\checkx^{\pullb}: \gphys \rightarrow G_{\text{check}}\;,
\end{equation}
where
\begin{equation}
\begin{gathered}
G_{\text{check}}\coloneqq \zz_2^{M^{(0)}} \times \zz_4^{M^{(1)}} \times \zz_8^{M^{(2)}}\\
\simeq\zz_2^{\nrxch(\nrxch-1)(\nrxch-2)/6} \times \zz_4^{\nrxch(\nrxch-1)/2} \times \zz_8^{\nrxch}\;.
\end{gathered}
\end{equation}
Here, $M^{(0)}$ is the set of all triples of $X$-checks, $M^{(1)}$ is the set of all pairs, and $M^{(2)}$ is the set of all individual $X$ checks.
So the subset of coefficients $\mathbf c$ giving rise to logical gates is determined by the simple equation
\footnote{Note that even the non-simplified condition in Eq.~\eqref{eq:cohomology_invariant} yields a similar equation,
\[\Big((\idop\mid \checkx)^{\pullb}-\idop\Big) \mathbf c=0\;,\]
where $(\idop\mid \checkx)$ denotes horizontal stacking of matrices.
So this again just involves finding the kernel of a group homomorphism.}
\begin{equation}
\label{eq:logical_condition_final}
\checkx^{\pullb} \mathbf c=0\;.
\end{equation}
Furthermore, we find that $\checkx^{\pullb}$ defines a group homomorphism,
\begin{equation}
\begin{gathered}
S^{\nrxch,\bm M}_{\checkx^{\pullb}(\mathbf c_0+\mathbf c_1)}
\overset{\eqref{eq:pullback_def}}{=} S^{n,\mathbf N}_{\mathbf c_0+\mathbf c_1} \circ \checkx\\
\overset{\eqref{eq:s_linear}}{=} S^{n,\mathbf N}_{\mathbf c_0} \circ \checkx + S^{n,\mathbf N}_{\mathbf c_1} \circ \checkx
\overset{\eqref{eq:pullback_def},\eqref{eq:s_linear}}{=} S^{\nrxch,\bm M}_{\checkx^{\pullb}\mathbf c_0+\checkx^{\pullb}\mathbf c_1}\;.
\end{gathered}
\end{equation}
So the solutions to Eq.~\eqref{eq:logical_condition_final} form a subgroup of $\gphys$, namely the kernel of the abelian group homomorphism $\checkx^{\pullb}$.
Finding logical gates can thus be reduced to two computational tasks.
The first task is to efficiently construct the coefficient matrix of $\checkx^{\pullb}$, which we do in Section~\ref{sec:constructing_coefficients}.
The second task is to efficiently compute the kernel of $\checkx^{\pullb}$, which we do in Section~\ref{sec:2group_linalg}.

\subsection{Constructing coefficient matrix of \texorpdfstring{$\checkx^{\pullb}$}{A*}}
\label{sec:constructing_coefficients}
Next we show how to compute the coefficient matrix of the homomorphism
\begin{equation}
\begin{multlined}
\checkx^{\pullb}: \gphys=\zz_2^{N^{(0)}}\times \zz_4^{N^{(1)}}\times \zz_8^{N^{(2)}}\\
\rightarrow G_{\text{check}}=\zz_2^{M^{(0)}}\times \zz_4^{M^{(1)}}\times \zz_8^{M^{(2)}}\;.
\end{multlined}
\end{equation}
Every homomorphism $X:\zz_a\rightarrow \zz_b$ is of the form
\begin{equation}
X(g)=c\frac{b}{\gcd(a,b)} g\;,
\end{equation}
where $c=\frac{\gcd(a,b)}{b} X(1)\in\zz_{\gcd(a,b)}$.
Every homomorphism $X:\prod_i G_i\rightarrow \prod_j H_j$ is given by
\begin{equation}
X(g)_j = \sum_i X_{ji}(g_i)
\end{equation}
where $X_{ji}: G_i\rightarrow H_j$ is the homomorphism defined by $X_{ji}(g)= X(g|_i)_j$, and
\begin{equation}
\label{eq:argument_position_notation}
g|_i \coloneqq (\underbrace{0,\ldots,0}_{0\ldots i-1},\underbrace{g}_i,\underbrace{0,\ldots,0}_{i+1\ldots n-1})\in G\;,\qquad g\in G_i\;.
\end{equation}
Taken together, $\checkx^{\pullb}$ is specified by a $(|M^{(0)}|+|M^{(1)}|+|M^{(2)}|)\times (|N^{(0)}|+|N^{(1)}|+|N^{(2)}|)$ coefficient matrix
\begin{equation}
\label{eq:coefficient_block_shape}
\begin{gathered}
(\checkx^{\pullb})_{(q,\mathbf x),(l,\mathbf i)}
\in \zz_{\gcd(2^{l+1},2^{q+1})}
= \zz_{2^{\min(l+1,q+1)}}\\
=
\begin{tabular}{r|lll}
\diagbox{$\scriptstyle{l}$}{$\scriptstyle{q}$} & $0$ & $1$ & $2$\\
\hline
$0$ & $\zz_2$ & $\zz_2$ &$\zz_2$\\
$1$ & $\zz_2$&$\zz_4$&$\zz_4$\\
$2$ & $\zz_2$&$\zz_4$&$\zz_8$
\end{tabular}
\;.
\end{gathered}
\end{equation}
Let $\delta^{(l)}_{\mathbf i}$ denote a canonical generator of $\gphys$, where $\mathbf i\in N^{(l)}$ with $l\in \{0,1,2\}$.
Then the coefficient we would like to compute is given by
\begin{equation}
\begin{gathered}
(\checkx^{\pullb})_{(q,\mathbf x),(l,\mathbf i)}
=
\frac{\gcd(2^{q+1},2^{l+1})}{2^{q+1}}
\big(\checkx^{\pullb} \delta^{(l)}_{\mathbf i}\big)^{(q)}_{\mathbf x}\\
=
2^{\min(q,l)-q}
\big(\checkx^{\pullb} \delta^{(l)}_{\mathbf i}\big)^{(q)}_{\mathbf x}
\;.
\end{gathered}
\end{equation}
for $q\in \{0,1,2\}$ and $\mathbf x\in M^{(q)}$.
Central to computing these coefficients is the following identity of the lift $\ovl\bullet: \zz_2\rightarrow\zz$:
\begin{equation}
\ovl{a+b}=\ovl a+\ovl b - 2\ovl a\ovl b\;,
\end{equation}
where the addition on the left-hand side is $\zz_2$ group multiplication.
Iterating this identity, we obtain
\begin{equation}
\label{eq:lift_expansion}
\begin{gathered}
\ovl{\sum_{0\leq i<n} a_i}
= \ovl a_0 + \ovl{\sum_{1\leq i<n} a_i} - 2\ovl a_0 \cdot \ovl{\sum_{1\leq i<n} a_i}=\ldots\\
= \sum_{0\leq i<n} \ovl a_i - 2\cdot \sum_{\mathclap{0\leq i<j<n}} \ovl a_i\ovl a_j + 4\cdot \sum_{\mathclap{0\leq i<j<k<n}} \ovl a_i \ovl a_j \ovl a_k - 8\cdot \ldots\;.
\end{gathered}
\end{equation}
Now, for $\delta^{(2)}_{\mathbf i}$ with $\mathbf i = i$, we compute
\begin{equation}
\begin{gathered}
S_{\checkx^{\pullb} \delta^{(2)}_i} (a)
= (S_{\delta^{(2)}_i} \circ \checkx)(a)
= S_{\delta^{(2)}_i}(\checkx(a))\\
= \frac18 \ovl{(\checkx(a))}_i
= \frac18 \ovl{\sum_x \checkx_{ix} a_x}\\
\overset{\eqref{eq:lift_expansion}}{=} \frac18 \sum_{0\leq x<\nrxch} \ovl{\checkx}_{ix} \ovl a_x - \frac14 \cdot \sum_{0\leq x<y<\nrxch} \ovl{\checkx}_{ix}\ovl{\checkx}_{iy} \ovl a_x\ovl a_y\\
 + \frac12 \cdot \sum_{0\leq x<y<z<\nrxch} \ovl{\checkx}_{ix}\ovl{\checkx}_{iy}\ovl{\checkx}_{iz} \ovl a_x \ovl a_y \ovl a_z\;.
\end{gathered}
\end{equation}
We confirm that $S_{\checkx^{\pullb} \delta^{(2)}_i}$ is indeed a third-order function, and read off the coefficients:
\begin{equation}
\begin{gathered}
(\checkx^{\pullb})_{(2,x),(2,i)} = (\checkx^{\pullb} \delta^{(2)}_i)^{(2)}_x = \ovl\checkx_{ix}\;,\\
(\checkx^{\pullb})_{(1,xy),(2,i)} = (\checkx^{\pullb} \delta^{(2)}_i)^{(1)}_{xy}= -\ovl\checkx_{ix}\ovl\checkx_{iy}\;,\\
(\checkx^{\pullb})_{(0,xyz),(2,i)} = (\checkx^{\pullb} \delta^{(2)}_i)^{(0)}_{xyz}= \ovl\checkx_{ix}\ovl\checkx_{iy}\ovl\checkx_{iz}\;.
\end{gathered}
\end{equation}
For $\delta^{(1)}_{\mathbf i}$ with $\mathbf i = ij$, we find
\begin{widetext}
\begin{equation}
\begin{gathered}
S_{\checkx^{\pullb} \delta^{(1)}_{ij}} (a)
= (S_{\delta^{(1)}_{ij}} \circ \checkx)(a)
= S_{\delta^{(1)}_{ij}}(\checkx(a))
= \frac14 \ovl{(\checkx(a))}_i \ovl{(\checkx(a))}_j
= \frac14 \Big(\ovl{\sum_x\checkx_{ix}a_x}\Big) \Big(\ovl{\sum_y \checkx_{jy}a_y}\Big)\\
\overset{\eqref{eq:lift_expansion}}{=} 
\Big(\frac14 \sum_x \ovl\checkx_{ix}\ovl a_x - \frac12 \sum_{x<z}\ovl\checkx_{ix}\ovl a_x \ovl\checkx_{iz}\ovl a_z\Big) \Big(\ovl{\sum_y \checkx_{jy}a_y}\Big)\\
\overset{\eqref{eq:lift_expansion}}{=} 
\Big(\sum_x \ovl\checkx_{ix}\ovl a_x\Big) \Big(\frac14 \sum_y \ovl\checkx_{jy}\ovl a_y - \frac12 \sum_{y<z} \ovl\checkx_{jy}\ovl a_y\ovl\checkx_{jz}\ovl a_z\Big) - \frac12 \Big(\sum_{x<z}\ovl\checkx_{ix}\ovl a_x \ovl\checkx_{iz}\ovl a_z\Big) \Big(\sum_y \ovl\checkx_{jy}\ovl a_y\Big)\\
=\frac14 \sum_{x,y} \ovl\checkx_{ix} \ovl\checkx_{jy} \ovl a_x\ovl a_y - \frac12 \sum_{x,y<z} \ovl\checkx_{ix} \ovl\checkx_{jy}\ovl\checkx_{jz}\ovl a_x\ovl a_y\ovl a_z-\frac12 \sum_{x<z,y} \ovl\checkx_{ix} \ovl\checkx_{iz}\ovl\checkx_{jy}\ovl a_x \ovl a_y \ovl a_z\\
=\frac18 \sum_{x} 2\ovl\checkx_{ix} \ovl\checkx_{jx} \ovl a_x\\
+\frac14 \sum_{x<y}\Big(
\ovl\checkx_{ix} \ovl\checkx_{jy}
+\ovl\checkx_{jx} \ovl\checkx_{iy}
+2 \ovl\checkx_{ix} \ovl\checkx_{jx}\ovl\checkx_{jy}
+2 \ovl\checkx_{iy} \ovl\checkx_{jx}\ovl\checkx_{jy}
+2 \ovl\checkx_{ix} \ovl\checkx_{iy}\ovl\checkx_{jx}
+2 \ovl\checkx_{ix} \ovl\checkx_{iy}\ovl\checkx_{jy}
\Big) \ovl a_x\ovl a_y\\
+ \frac12 \sum_{x<y<z} \Big(\ovl\checkx_{ix} \ovl\checkx_{jy}\ovl\checkx_{jz}+ \ovl\checkx_{iy}\ovl\checkx_{jx}\ovl\checkx_{jz}+\ovl\checkx_{iz} \ovl\checkx_{jx}\ovl\checkx_{jy} +\ovl\checkx_{iy} \ovl\checkx_{iz}\ovl\checkx_{jx}+\ovl\checkx_{ix} \ovl\checkx_{iz}\ovl\checkx_{jy}+\ovl\checkx_{ix} \ovl\checkx_{iy}\ovl\checkx_{jz}\Big) \ovl a_x \ovl a_y \ovl a_z\\
\;.
\end{gathered}
\end{equation}
\end{widetext}
In the last step, we have reorganized the sums.
For example, the sum over $x$ and $y$ can be divided into sums over $x<y$, $x=y$, and $x>y$.
Or, the sum over $x,y<z$ can be divided into sums over $x<y<z$, $x=y<z$, $y<x<z$, $y<z=x$, and $y<z<x$.
After dividing them, the different sums are then recombined.
We read off
\begin{equation}
\begin{gathered}
(\checkx^{\pullb})_{(2,x),(1,ij)} = \frac12 (\checkx^{\pullb} \delta^{(1)}_{ij})^{(2)}_x = \ovl\checkx_{ix} \ovl\checkx_{jx}\;,\\
\begin{multlined}
(\checkx^{\pullb})_{(1,xy),(1,ij)} =(\checkx^{\pullb} \delta^{(1)}_{ij})^{(1)}_{xy}\\=
\ovl\checkx_{ix} \ovl\checkx_{jy}
+\ovl\checkx_{jx} \ovl\checkx_{iy}
+2 \ovl\checkx_{ix} \ovl\checkx_{jx}\ovl\checkx_{jy}\\
+2 \ovl\checkx_{iy} \ovl\checkx_{jx}\ovl\checkx_{jy}
+2 \ovl\checkx_{ix} \ovl\checkx_{iy}\ovl\checkx_{jx}
+2 \ovl\checkx_{ix} \ovl\checkx_{iy}\ovl\checkx_{jy}\;,
\end{multlined}
\\
\begin{multlined}
(\checkx^{\pullb})_{(0,xyz),(1,ij)} =
(\checkx^{\pullb} \delta^{(1)}_{ij})^{(0)}_{xyz}\\
= \ovl\checkx_{ix} \ovl\checkx_{jy}\ovl\checkx_{jz}+ \ovl\checkx_{iy}\ovl\checkx_{jx}\ovl\checkx_{jz}+\ovl\checkx_{iz} \ovl\checkx_{jx}\ovl\checkx_{jy}\\ +\ovl\checkx_{iy} \ovl\checkx_{iz}\ovl\checkx_{jx}+\ovl\checkx_{ix} \ovl\checkx_{iz}\ovl\checkx_{jy}+\ovl\checkx_{ix} \ovl\checkx_{iy}\ovl\checkx_{jz}\;.
\end{multlined}
\end{gathered}
\end{equation}
Finally, for $\delta^{(0)}_{\mathbf i}$ with $\mathbf i=ijk$, we find
\begin{widetext}
\begin{equation}
\begin{gathered}
S_{\checkx^{\pullb} \delta^{(0)}_{ijk}} (a)
= (S_{\delta^{(0)}_{ijk}} \circ \checkx)(a)
= S_{\delta^{(0)}_{ijk}}(\checkx(a))
= \frac12 \ovl{(\checkx(a))}_i \ovl{(\checkx(a))}_j \ovl{(\checkx(a))}_k
= \frac12 \Big(\ovl{\sum_x\checkx_{ix}a_x}\Big) \Big(\ovl{\sum_y \checkx_{jy}a_y}\Big) \Big(\ovl{\sum_z \checkx_{kz}a_z}\Big)\\
= \frac12 \Big(\sum_x\ovl\checkx_{ix}\ovl a_x\Big) \Big(\sum_y \ovl\checkx_{jy}\ovl a_y\Big) \Big(\sum_z \ovl\checkx_{kz}\ovl a_z\Big)
= \frac12 \sum_{x,y,z} \ovl\checkx_{ix} \ovl\checkx_{jy} \ovl\checkx_{kz}\ovl a_x \ovl a_y \ovl a_z\\
= \frac18 \sum_x 4 \ovl\checkx_{ix} \ovl\checkx_{jx} \ovl\checkx_{kx}\ovl a_x\\
+ \frac14  \sum_{x<y} 2\Big(
\ovl\checkx_{ix} \ovl\checkx_{jx} \ovl\checkx_{ky}
+\ovl\checkx_{ix} \ovl\checkx_{jy} \ovl\checkx_{kx}
+\ovl\checkx_{ix} \ovl\checkx_{jy} \ovl\checkx_{ky}
+\ovl\checkx_{iy} \ovl\checkx_{jy} \ovl\checkx_{kx}
+\ovl\checkx_{iy} \ovl\checkx_{jx} \ovl\checkx_{ky}
+\ovl\checkx_{iy} \ovl\checkx_{jx} \ovl\checkx_{kx}
\Big) \ovl a_x \ovl a_y\\
+ \frac12 \sum_{x<y<z} \big(\ovl\checkx_{ix} \ovl\checkx_{jy} \ovl\checkx_{kz} + \ovl\checkx_{ix} \ovl\checkx_{jz} \ovl\checkx_{ky}+\ovl\checkx_{iy} \ovl\checkx_{jx} \ovl\checkx_{kz}+\ovl\checkx_{iy} \ovl\checkx_{jz} \ovl\checkx_{kx}+\ovl\checkx_{iz} \ovl\checkx_{jx} \ovl\checkx_{ky}+\ovl\checkx_{iz} \ovl\checkx_{jy} \ovl\checkx_{kx}\Big) \ovl a_x \ovl a_y \ovl a_z
\;.
\end{gathered}
\end{equation}
\end{widetext}
Again, we have divided the sum over independent $x,y,z$ into cases like $x=y=z$, $x=y<z$, or $x<z<y$.
We read off
\begin{equation}
\begin{gathered}
(\checkx^{\pullb})_{(2,x),(0,ijk)} =\frac14 (\checkx^{\pullb} \delta^{(0)}_{ijk})^{(2)}_x = \ovl\checkx_{ix} \ovl\checkx_{jx} \ovl\checkx_{kx}\;,\\
\begin{multlined}
(\checkx^{\pullb})_{(1,xy),(0,ijk)} =\frac12 (\checkx^{\pullb} \delta^{(0)}_{ijk})^{(1)}_{xy}\\
=\ovl\checkx_{ix} \ovl\checkx_{jx} \ovl\checkx_{ky}
+\ovl\checkx_{ix} \ovl\checkx_{jy} \ovl\checkx_{kx}
+\ovl\checkx_{ix} \ovl\checkx_{jy} \ovl\checkx_{ky}\\
+\ovl\checkx_{iy} \ovl\checkx_{jy} \ovl\checkx_{kx}
+\ovl\checkx_{iy} \ovl\checkx_{jx} \ovl\checkx_{ky}
+\ovl\checkx_{iy} \ovl\checkx_{jx} \ovl\checkx_{kx}\;,
\end{multlined}
\\
\begin{multlined}
(\checkx^{\pullb})_{(0,xyz),(0,ijk)} =(\checkx^{\pullb} \delta^{(0)}_{ijk})^{(0)}_{xyz}\\
= \ovl\checkx_{ix} \ovl\checkx_{jy} \ovl\checkx_{kz} + \ovl\checkx_{ix} \ovl\checkx_{jz} \ovl\checkx_{ky}+\ovl\checkx_{iy} \ovl\checkx_{jx} \ovl\checkx_{kz}\\+\ovl\checkx_{iy} \ovl\checkx_{jz} \ovl\checkx_{kx}+\ovl\checkx_{iz} \ovl\checkx_{jx} \ovl\checkx_{ky}+\ovl\checkx_{iz} \ovl\checkx_{jy} \ovl\checkx_{kx}\;.
\end{multlined}
\end{gathered}
\end{equation}

\subsection{Computing the kernel of \texorpdfstring{$\checkx^{\pullb}$}{A*}}
\label{sec:2group_linalg}
In this section, we show how to efficiently compute the kernel of a group homomorphism
\begin{equation}
\label{eq:xhom}
X: \zz_2^{n_0}\times \zz_4^{n_1}\times \zz_8^{n_2} \rightarrow \zz_2^{m_0}\times \zz_4^{m_1}\times \zz_8^{m_2}\;,
\end{equation}
given by a coefficient matrix as discussed around Eq.~\eqref{eq:coefficient_block_shape}.
We will apply this to $X=\checkx^{\pullb}$, and switch to the letter $X$ merely for notational compactness.
More precisely, we compute a \emph{kernel isomorphism}
\begin{equation}
K: \zz_2^{k_0}\times \zz_4^{k_1}\times \zz_8^{k_2} \rightarrow \zz_2^{n_0}\times \zz_4^{n_1}\times \zz_8^{n_2}\;,
\end{equation}
which is an injective homomorphism such that $XK=0$.
The output of the algorithm consists of (1) the three numbers $(k_0,k_1,k_2)$ and (2) the coefficient matrix of $K$.
The generalization to arbitrary finite abelian 2-groups (groups consisting of factors of the form $\zz_{2^l}$) is straightforward and discussed in Section~\ref{sec:higher_clifford}.

There are various methods for computing the kernel isomorphism $K$.
A common algorithm is to use ``slack variables'' to turn the finite-group equation into integer equations, and then find the kernel using the \emph{Smith normal form}.
Here, we propose a different method that we call \emph{filtration}, which uses only binary ($\zz_2$) matrix operations as subroutines.
Filtration has the advantage that it is much faster than the Smith normal form:
$\zz_2$ arithmetic can be sped up by bit-packing, whereas the entries in integer arithmetic can grow very large.
We do not know whether this method is described anywhere in detail in the literature, though it is somewhat related to the notion of \emph{Hensel lifting}~\cite{Hartung2010, Dixon1982}.
Roughly speaking, filtration makes use of the fact that
\begin{equation}
\begin{gathered}
\ker(X)\subset \ker(X\mmod 4)\subset \ker(X\mmod 2)\\\subset \zz_2^{n_0}\times \zz_4^{n_1}\times \zz_8^{n_2}\;.
\end{gathered}
\end{equation}
The embedding of each kernel into the next is determined by the kernel of a homomorphism whose codomain is of the form $\zz_2^m$, which is easier to compute.

\myparagraph{Notation and simple subroutines}
We start by fixing some notation, and reviewing some elementary subroutines for binary ($\zz_2$) linear algebra.
For $l<k$, let $\mmod_{2^l}$ denote the homomorphism
\begin{equation}
\mmod_{2^l}: \zz_{2^k}\rightarrow \zz_{2^l}\;,\quad \mmod_{2^l}(x) = x\mod 2^l\;.
\end{equation}
Specifically we will use $\mmod_2$ and $\mmod_4$.
We will denote the composition of group homomorphisms by simple concatenation, omitting the usual $\circ$ symbol.
To specify a homomorphism in terms of its $\zz_2$, $\zz_4$ or $\zz_8$ coefficients, we will arrange these coefficients into a (block) matrix with straight delimiters, such as
\begin{equation}
\coeffs{a&b\\c&d}\;.
\end{equation}
Note that we will talk about the coefficient matrices and corresponding homomorphisms somewhat interchangeably.
Similar to Section~\ref{sec:constructing_coefficients}, we will use $\ovl\bullet$ to denote the canonical lift, embedding $[0,2^i-1]$ into $[0,\ldots,2^j-1]$, where $i<j$ depend on the context.
Also, we let $\cdot 2$ denote the obvious homomorphism from $\zz_{2^k}\rightarrow \zz_{2^{k+1}}$.

We now describe some elementary subroutines.
The first is computing a \emph{$\zz_2$ staggered kernel isomorphism} $K$ of a homomorphism $X: \zz_2^{n_0+n_1+n_2}\rightarrow \zz_2^m$.
This is an injective homomorphism $K: \zz_2^{k_0+k_1+k_2}\rightarrow\zz_2^{n_0+n_1+n_2}$ such that (1) $XK=0$ and (2) written as a $3\times 3$ block matrix, we have
\begin{equation}
K=
\coeffs{
K_{00} & K_{01} & K_{02}\\
0 & K_{11} & K_{12}\\
0 & 0 & K_{22}
}
\;,
\end{equation}
and (3) $K_{00}$, $K_{11}$, and $K_{22}$ are all injective.
$K$ can be constructed rather directly from the \emph{row-reduced echelon form (RREF)} of $X$.
The second subroutine is the \emph{$\zz_2$ image completion} $B$ of an injective homomorphism $X: \zz_2^n\rightarrow \zz_2^m$.
This is an injective matrix $B$ such that the block matrix $\mpm{X&B}$ is bijective.
In fact, we can choose the columns of $B$ to consist of the computational basis vectors that are linearly independent from the columns of $X$ (and the previously chosen basis vectors).
These basis vectors are just the pivot columns in the second block of the RREF of $\mpm{X&\mathbb 1}$.

\myparagraph{$\zz_2^\bullet\times\zz_4^\bullet\times \zz_8^\bullet\rightarrow \zz_2^\bullet$ kernel}
The kernel algorithm makes use of the following major subroutine, which computes a kernel isomorphism of a homomorphism
\begin{equation}
X: \zz_2^{n_0}\times \zz_4^{n_1}\times \zz_8^{n_2} \rightarrow \zz_2^m\;,
\end{equation}
determined by a block coefficient matrix
\begin{equation}
X =
\coeffs{X_0&X_1&X_2}
\end{equation}
with all binary entries.
Any such homomorphism is given by $X=Y \mmod_2$ for some $Y:\zz_2^{n_0+n_1+n_2}\rightarrow\zz_2^m$, which has the same coefficient matrix as $X$.
The kernel of $X=Y\mmod_2$ is formed by all elements $k_x\in \zz_2^{n_0}\times \zz_4^{n_1}\times \zz_8^{n_2}$ such that $\mmod_2k_x\in \ker(Y)$.
Thus, the kernel is spanned by (1) a subgroup $\img(K_{\text{lift}})$ such that $\img(\mmod_2 K_{\text{lift}})=\ker(Y)$ (where $K_{\text{lift}}$ is some kernel homomorphism, $XK_{\text{lift}}=0$) and (2) the subgroup $\ker(\mmod_2)=\img(K_{\text{triv}})$, where
\begin{equation}
\label{eq:trivial_generators}
\begin{gathered}
K_{\text{triv}}=
\begin{pmatrix}0&0\\\cdot 2&0\\0&\cdot 2\end{pmatrix}
=\coeffs{0&0\\\mathbb 1&0\\0&\mathbb 1}:\\
\zz_2^{n_1}\times \zz_4^{n_2}\rightarrow \zz_2^{n_0}\times \zz_4^{n_1}\times \zz_8^{n_2}\;.
\end{gathered}
\end{equation}
Since we want an injective kernel isomorphism $K$, we need to (1) define $K_{\text{lift}}$ such that it is injective, and (2) restrict $K_{\text{triv}}$ to the generators that are independent of those of $K_{\text{lift}}$, resulting in $\widetilde K_{\text{triv}}$.
The kernel isomorphism is then given by $K=\begin{pmatrix}K_{\text{lift}} & \widetilde K_{\text{triv}}\end{pmatrix}$.

To define $K_{\text{lift}}$, we compute a $\zz_2$ staggered kernel isomorphism $Z$ for $Y=\coeffs{X_0&X_1&X_2}$,
\begin{equation}
\label{eq:staggered_kernel_for_lift}
Z=
\coeffs{Z_{00}&Z_{01}&Z_{02}\\0&Z_{11}&Z_{12}\\0&0&Z_{22}}
: \zz_2^{h_0+h_1+h_2}\rightarrow \zz_2^{n_0+n_1+n_2}
\;.
\end{equation}
Then we choose
\begin{equation}
\label{eq:lift_generators}
\begin{multlined}
K_{\text{lift}}
=
\coeffs{Z_{00}&Z_{01}&Z_{02}\\0&\ovl Z_{11}&\ovl Z_{12}\\0&0&\ovl Z_{22}}:\\
\zz_2^{h_0}\times \zz_4^{h_1}\times \zz_8^{h_2}\rightarrow \zz_2^{n_0}\times \zz_4^{n_1}\times \zz_8^{n_2}
\;.
\end{multlined}
\end{equation}
Here, the lift is from $[0,1]$ to $[0,1,2,3]$ for $Z_{11}$ and $Z_{12}$, and from $[0,1]$ to $[0,\ldots,7]$ for $Z_{22}$.
As this lift is a right inverse function of the $\mmod_2$ homomorphism, this choice indeed satisfies $\img(\mmod_2K_{\text{lift}})=\img(Z)=\ker(Y)$.
Further, $Z$ is constructed such that $Z_{00}$, $Z_{11}$, and $Z_{22}$ are all injective.
Thus, the first column of $K_{\text{lift}}$ consists of independent order-2 generators, the second column consists of independent order-4 generators, and the third column consists of independent order-8 generators.
In other words, $K_{\text{lift}}$ is injective.

Next, we construct $\widetilde K_{\text{triv}}$ by choosing only the generators in $K_{\text{triv}}$ that are linearly independent from those in $\img(K_{\text{lift}})$.
In other words, $\img(K_{\text{lift}})$ and $\img(K_{\text{triv}})$ overlap, and we need to remove the overlap from $K_{\text{triv}}$.
Since the order-8 generators $x$ forming the last column block of $K_{\text{lift}}$ are independent, the only order-4 elements they span are of the form $2\cdot x$.
Note that the kernel-isomorphism coefficients of $2\cdot x$ are related to these of $x$ as
\begin{equation}
\coeffs{(2\cdot x)_0\\(2\cdot x)_1\\(2\cdot x)_2}
= \coeffs{0\\2 x_1\\x_2}\;,
\end{equation}
as the coefficient group changes from $\zz_8\rightarrow\zz_4$ in the third block row, and $\cdot 2$ maps everything to zero in the first block row with $\zz_2$ coefficients.
Since the order-4 generators $x$ forming the middle column block of $K_{\text{lift}}$ are independent, the only order-2 elements they span are of the form $2\cdot x$.
Taken together, the following linear combinations of generators of $\img(K_{\text{triv}})=\ker(\mmod_2)\simeq \zz_2^{n_1}\times \zz_4^{n_2}$ are already in $\img(K_{\text{lift}})$,
\begin{equation}
\label{eq:generators_to_remove}
\img
\coeffs{Z_{11}&Z_{12}\\0&\ovl Z_{22}}:
\zz_2^{h_1}\times \zz_4^{h_2}\rightarrow \zz_2^{n_1}\times \zz_4^{n_2}
\;.
\end{equation}
Now, let $W_1$ be a $\zz_2$ image completion of $Z_{11}$, and $W_2$ a $\zz_2$ image completion of $Z_{22}$.
Then we can set
\begin{equation}
\label{eq:independent_trivial_order4_gen}
\widetilde K_{\text{triv}}=
 K_{\text{triv}}\circ
\coeffs{W_1&0\\0&\ovl W_2}\;.
\end{equation}
By definition, $\img\coeffs{Z_{22}&W_2}$ is an isomorphism of $\zz_2^{n_2}$ groups, so $\img\coeffs{\ovl Z_{22}&\ovl W_2}$ is an isomorphism of $\zz_4^{n_2}$ groups.
Similarly, by definition, $\img\coeffs{Z_{11}&W_1}$ is an isomorphism of $\zz_2^{n_1}$ groups.
So the horizontal stack of Eq.~\eqref{eq:generators_to_remove} and $\widetilde K_{\text{triv}}$,
\begin{equation}
\left|\begin{array}{cc|cc}Z_{11}&W_1&Z_{12}&0\\0&0&\ovl Z_{22}&\ovl W_2\end{array}\right|\;,
\end{equation}
is an isomorphism of $\zz_2^{n_1}\times \zz_4^{n_2}$ groups.
Finally, combining $K_{\text{lift}}$ and $\widetilde K_{\text{triv}}$ into one matrix yields
\begin{equation}
K=
\left|
\begin{array}{cc|cc|c}
Z_{00}&0&Z_{01}&0&Z_{02}\\0&W_1&\ovl Z_{11}&0&\ovl Z_{12}\\0&0&0&\ovl W_2&\ovl Z_{22}
\end{array}
\right|\;.
\end{equation}
We summarize the subroutine in Algorithm~\ref{alg:z248_z2_kernel}.

\begin{figure}
\begin{algorithm}[H]
\caption{Kernel isomorphism $K$ of $X : \mathbb{Z}_2^{n_0} \times \mathbb{Z}_4^{n_1} \times \mathbb{Z}_8^{n_2} \to \mathbb{Z}_2^m$}
\label{alg:z248_z2_kernel}
\begin{algorithmic}[1]
\Require Binary coefficient matrix $\coeffs{X_0&X_1&X_2}$ of $X$
\Ensure $3\times 3$ block coefficient matrix of $K$
\State Compute a $\zz_2$ staggered kernel isomorphism
\[
\coeffs{Z_{00}&Z_{01}&Z_{02}\\0&Z_{11}&Z_{12}\\0&0&Z_{22}}
\]
of $\coeffs{X_0&X_1&X_2}$.
\State Compute the $\zz_2$ image completion $W_2$ of $Z_{22}$
\State Compute the $\zz_2$ image completion $W_1$ of $Z_{11}$
\State Return
\[
K=
\left|
\begin{array}{cc|cc|c}
Z_{00}&0&Z_{01}&0&Z_{02}\\0&W_1&\ovl Z_{11}&0&\ovl Z_{12}\\0&0&0&\ovl W_2&\ovl Z_{22}
\end{array}
\right|
\]
\end{algorithmic}
\end{algorithm}
\end{figure}

\myparagraph{Main kernel algorithm}
Given Algorithm~\ref{alg:z248_z2_kernel} as a subroutine, we can now describe the full algorithm finding a kernel isomorphism $K$ of $X$.
We start by using Algorithm~\ref{alg:z248_z2_kernel} to construct a kernel isomorphism $K_0=L_0$ of the homomorphism
\begin{equation}
\mmod_2X: \zz_2^{n_0}\times \zz_4^{n_1}\times \zz_8^{n_2} \rightarrow \zz_2^{m_0+m_1+m_2}\;.
\end{equation}
Since $\mmod_2X K_0=0$, the image of
\begin{equation}
\mmod_4XK_0: \zz_2^{l_0}\times \zz_4^{l_1}\times \zz_8^{l_2} \rightarrow \zz_2^{m_0}\times \zz_4^{m_1+m_2}
\end{equation}
only consists of even elements, and we can divide it by 2,
\begin{equation}
\frac12 \mmod_4XK_0: \zz_2^{l_0}\times \zz_4^{l_1}\times \zz_8^{l_2} \rightarrow \zz_2^{m_1+m_2}\;.
\end{equation}
We then use Algorithm~\ref{alg:z248_z2_kernel} again to find the kernel isomorphism $L_1$ of $\frac12 \mmod_4 XK_0$.
The kernel of $\mmod_4X$ is thus given by $K_1=K_0L_1$.
Since $\mmod_4XK_1=0$, the image of $XK_1$ is restricted to the subgroup of elements that are multiples of 4, so it can be divided by 4, yielding a homomorphism
\begin{equation}
\frac14 XK_1: \zz_2^{l_0}\times \zz_4^{l_1}\times \zz_8^{l_2} \rightarrow \zz_2^{m_2}\;.
\end{equation}
We use Algorithm~\ref{alg:z248_z2_kernel} a final time to find the kernel isomorphism $L_2$ of $\frac14 XK_1$.
The kernel isomorphism of $X$ is then given by $K=K_2=K_1L_2$.
The algorithm is summarized in Algorithm~\ref{alg:full_kernel}.

\begin{figure}
\begin{algorithm}[H]
\caption{Kernel isomorphism $K$ of $X : \mathbb{Z}_2^{n_0} \times \mathbb{Z}_4^{n_1} \times \mathbb{Z}_8^{n_2} \to \mathbb{Z}_2^{m_0}\times \mathbb{Z}_4^{m_1}\times \mathbb{Z}_8^{m_2}$}
\label{alg:full_kernel}
\begin{algorithmic}[1]
\Require $3\times 3$ block coefficient matrix of $X$
\Ensure $3\times 3$ block coefficient matrix of $K$
\State Use Algorithm~\ref{alg:z248_z2_kernel} to compute a kernel isomorphism $L_0$ of
\[
\mmod_2X
=
\coeffs{
X_{00}&X_{01}&X_{02}\\
0&X_{11}\mmod 2 & X_{12}\mmod 2\\
0&0&X_{22}\mmod 2
}
\]
\State Set $K_0=L_0$
\State Use Algorithm~\ref{alg:z248_z2_kernel} to compute a kernel isomorphism $L_1$ of $\frac12 \mmod_4 XK_0$
\State Set $K_1=K_0L_1$
\State Use Algorithm~\ref{alg:z248_z2_kernel} to compute a kernel isomorphism $L_2$ of $\frac14 XK_1$
\State Return $K=K_2=K_1L_2$
\end{algorithmic}
\end{algorithm}
\end{figure}

\myparagraph{Overall runtime}
We now briefly analyze the runtime scaling of the algorithm.
First, let us look at the size of the coefficient matrix of $\checkx^{\pullb}$, that is, the number of generators in $\gphys$ and $G_{\text{check}}$.
There are in principle $O(n^3)$ possible ansatz gates, corresponding to the $O(n^3)$ qubit triples where a $CCZ$ gate may act.
However, for a locality-preserving logical gate, every qubit should only be involved in $O(1)$ different $T$, $CS$, or $CCZ$ ansatz gates.
So in this case, the number of gate locations and generators of $\gphys$ is $O(n)$.
Next, we have formally defined $G_{\text{check}}$ to be fixed to $O(m^3)=O(n^3)$ generators.
However, the image of $\checkx^{\pullb}$ is restricted to a smaller support by locality.
The support can only contain a pair or triple of checks if there is a physical ansatz gate whose support overlaps with both or all three checks.
In a qLDPC code family the full interaction graph on the qubits including $X$ checks and physical ansatz gates should have constant bounded degree.
So after restricting $\checkx^{\pullb}$ to the generators in its support, there are $O(m)=O(n)$ such generators again.
The coefficient matrix for $\checkx^{\pullb}$ is thus a matrix of dimension $O(n)\times O(n)$.

Further, the coefficient matrix for $\checkx^{\pullb}$ is both row and column sparse, and specified by $O(n)$ coefficients.
A sparse representative of $\checkx^{\pullb}$ can be computed in $O(n)$ time.
The runtime of Algorithm~\ref{alg:full_kernel} depends on the implementation.
Using naive dense binary ($\zz_2$) linear algebra, the runtime is $O(n^3)$, as both RREF and matrix multiplication of $O(n)\times O(n)$-size matrices require $O(n^3)$ runtime.
Note that this runtime scaling is the same as for finding the code space dimension/logical basis of the CSS code using dense $\zz_2$ linear algebra.
In fact, as we briefly discuss in Section~\ref{sec:conclusion}, the problem of finding locality-preserving logical gates in a qLDPC code is equivalent to finding a logical basis of another qLDPC code, consisting of qudits and enlarged by some constant.
As we briefly discuss in Section~\ref{sec:faster_locality}, it may be possible to improve the (practical or asymptotic) runtime of the algorithms by making use of locality/sparsity for qLDPC codes.

\subsection{Implementation and examples}
A python implementation of the described methods can be found in the accompanying code repository Ref.~\cite{diagonal_gate_finder}.
The linear-algebra routines for abelian 2-groups are provided in a separate python package \texttt{twogroup-linalg}~\cite{twogroup_linalg}.
The implementation allows one to define an arbitrary set of diagonal ansatz gates at arbitrary levels of the Clifford hierarchy, and then provides efficient methods to find the group of all logical gates composed from these ansatz gates, or to find a physical implementation of a desired logical operation.
We also implement an efficient method for finding translation-invariant locality-preserving logical gates in translation-invariant CSS codes in any spatial dimension.
In this case, we implement two ways of determining the non-triviality of a translation-invariant gate:
Either we define a gate trivial if it is a sum over translates of locally supported logical gates, or we define it as trivial if it has trivial logical action on a given finite compactification of the code.
The first method is fully independent of the code size.
The second method scales with the size of the finite compactified code, but is generally much faster than finding non-translation-invariant gates built from the same ansatz gates.

For quotienting by the subgroup of trivial logical gates and finding physical gate representatives, we implement ``filtration''-based methods for the epi-mono decomposition of a homomorphism between abelian 2-groups, and for solving 2-group-based linear equations.
The latter algorithm is described in Appendix~\ref{sec:2group_solving}.

The current implementation uses dense binary ($\zz_2$) linear algebra.
For the $\zz_2$ RREF, we use the bit-packed rust library \texttt{bitgauss}~\cite{Kissinger2025}, but composition of homomorphisms is currently not bit-packed.
The method is fast for reasonably-sized codes, for example we can find transversal gates on a 2D color code with 800 qubits in 30 seconds on a standard laptop (for the non-translation-invariant case; in the translation-invariant case the runtime is well below a second).
However, adding more ansatz gate locations, in particular multi-qubit locations, does increase the runtime significantly.

In the accompanying jupyter notebook various examples are considered.
We confirm the well-known logical gates for instances of the 2D and 3D toric and color codes, fracton codes, and bivariate bicycle codes~\cite{Bravyi2023}.
As an application, we find the previously unknown third-order diagonal gate in the ``dual 3D color code''.
This is a code with qubits on the faces of a 3-colex, two independent $X$ stabilizers per volume, and one $Z$ stabilizer per vertex.
If we use this code for a measurement-based dimension-jump protocol, the resulting length-3 CSS fault complex is reversed compared to the usual color-code-based protocol.
The third-order gate we find implements the same logical action as the $T$ gate in the usual color code.

\section{Generalizations and improvements}
\label{sec:generalizations}
In this section, we discuss several generalizations of the methods in Section~\ref{sec:theory}.

\subsection{Spacetime local logical gates}
\label{sec:spacetime_logical}
Here, we show how our algorithm can also be used to efficiently find spacetime local logical gates.

\myparagraph{CSS circuits}
The spacetime generalization of a CSS code is a \emph{CSS circuit}, by which we mean a circuit consisting of $CX$ gates as well as $Z^{\otimes i}$ and $X^{\otimes i}$ measurements.
A CSS circuit can be turned into a tensor-network diagram consisting of two types of tensors, namely \emph{GHZ tensors},
\begin{equation}
\begin{tikzpicture}
\atoms{circ,small,all}{0/}
\draw (0)--++(0:0.5) (0)--++(90:0.5) (0)--++(180:0.5);
\node at (-90:0.2){$\ldots$};
\end{tikzpicture}
=
\ket{000\ldots}+\ket{111\ldots}\;,
\end{equation}
and \emph{sum tensors},
\begin{equation}
\begin{tikzpicture}
\atoms{circ,small}{0/}
\draw (0)--++(0:0.5) (0)--++(90:0.5) (0)--++(180:0.5);
\node at (-90:0.2){$\ldots$};
\end{tikzpicture}
=
\sum_{\substack{a,b,c,\ldots:\\a+b+c+\ldots = 0\mmod 2}} \ket{abc\ldots}\;.
\end{equation}
Such tensor networks have also been popularized under the name \emph{ZX calculus}~\cite{Coecke2017,Wetering2020,Kissinger2022}.
The tensor network is obtained by replacing circuit elements with tensor-network snippets as follows,
\begin{equation}
\label{eq:tensor_network_translation}
\begin{tabular}{ccccc}
$CX$ &$=$&
\begin{tikzpicture}
\atoms{circ,tiny,all}{0/}
\atoms{circ,small,cross}{1/p={0.8,0}}
\draw (0)--(1) (0)--++(-90:0.5) (0)--++(90:0.5) (1)--++(-90:0.5) (1)--++(90:0.5);
\end{tikzpicture}
&$\rightarrow$&
\begin{tikzpicture}
\atoms{circ,small,all}{0/}
\atoms{circ,small}{1/p={0.8,0}}
\draw (0)--(1) (0)--++(-90:0.5) (0)--++(90:0.5) (1)--++(-90:0.5) (1)--++(90:0.5);
\end{tikzpicture}
\;,\\[20pt]
$\text{meas.}Z^{\otimes i}$
&$=$&
\begin{tikzpicture}
\node[draw,rectangle,minimum width=2cm] (0) {$Z^{\otimes i}$};
\draw ([sx=-0.75]0.south)--++(-90:0.4) ([sx=-0.25]0.south)--++(-90:0.4) ([sx=0.25]0.south)--++(-90:0.4);
\draw ([sx=-0.75]0.north)--++(90:0.4) ([sx=-0.25]0.north)--++(90:0.4) ([sx=0.25]0.north)--++(90:0.4);
\node at (0.75,-0.4){$\ldots$};
\node at (0.75,0.4){$\ldots$};
\end{tikzpicture}
&$\rightarrow$&
\begin{tikzpicture}
\atoms{circ,small,all}{0/, 1/p={0.5,0}, 2/p={1,0}}
\atoms{circ,small}{x/p={0.75,0.5}}
\draw (x)--(0) (x)--(1) (x)--(2) (0)--++(-90:0.5) (0)--++(90:0.8) (1)--++(-90:0.5) (1)--++(90:0.8) (2)--++(-90:0.5) (2)--++(90:0.8) (x)edge[mark={three dots,a}]($(x)!0.5!(1.5,0)$);
\end{tikzpicture}
\;,\\[20pt]
$\text{meas.}X^{\otimes i}$
&$=$&
\begin{tikzpicture}
\node[draw,rectangle,minimum width=2cm] (0) {$X^{\otimes i}$};
\draw ([sx=-0.75]0.south)--++(-90:0.4) ([sx=-0.25]0.south)--++(-90:0.4) ([sx=0.25]0.south)--++(-90:0.4);
\draw ([sx=-0.75]0.north)--++(90:0.4) ([sx=-0.25]0.north)--++(90:0.4) ([sx=0.25]0.north)--++(90:0.4);
\node at (0.75,-0.4){$\ldots$};
\node at (0.75,0.4){$\ldots$};
\end{tikzpicture}
&$\rightarrow$&
\begin{tikzpicture}
\atoms{circ,small}{0/, 1/p={0.5,0}, 2/p={1,0}}
\atoms{circ,small,all}{x/p={0.75,0.5}}
\draw (x)--(0) (x)--(1) (x)--(2) (0)--++(-90:0.5) (0)--++(90:0.8) (1)--++(-90:0.5) (1)--++(90:0.8) (2)--++(-90:0.5) (2)--++(90:0.8) (x)edge[mark={three dots,a}]($(x)!0.5!(1.5,0)$);
\end{tikzpicture}
\;,
\end{tabular}
\end{equation}
where time in the picture goes upward.
The linear operator corresponding to the tensor network diagram for the CX gate is equal to the CX unitary.
The linear operator corresponding to the tensor network diagram for the $X$ or $Z$-type measurement is equal to the measurement projector for the $+1$ outcome.
\footnote{Here we are ignoring global normalization factors which have no physical meaning.}
In other words, the tensor network is the same as the circuit where all measurement outcomes are postselected to $+1$.

We now slightly deform the tensor-network diagram.
First, we fuse adjacent pairs of neighboring GHZ or sum tensors, for example,
\begin{equation}
\begin{tikzpicture}
\atoms{circ,all,small}{0/, 1/p={0.8,0}}
\draw (0)--(1) (0)--++(135:0.5) (0)--++(-135:0.5) (1)--++(90:0.5) (1)--++(0:0.5) (1)--++(-90:0.5);
\end{tikzpicture}
=
\begin{tikzpicture}
\atoms{circ,all,small}{0/}
\draw (0)--++(135:0.5) (0)--++(-135:0.5) (0)--++(90:0.5) (0)--++(0:0.5) (0)--++(-90:0.5);
\end{tikzpicture}
\;.
\end{equation}
This turns the tensor network into a full-empty bipartite graph.
Next, for every input and output wire of a CSS circuit that is adjacent to a GHZ tensor, we add a 2-index dummy sum tensor corresponding to an identity operator.
For example, for a GHZ tensor with two input indices at the bottom, we add sum tensors as
\begin{equation}
\begin{tikzpicture}
\atoms{circ,all,small}{0/}
\draw (0)edge[mark={three dots,a}]++(60:0.5) (0)edge[mark={three dots,a}]++(120:0.5) (0)--++(-60:0.8) (0)--++(-120:0.8);
\end{tikzpicture}
=
\begin{tikzpicture}
\atoms{circ,all,small}{0/}
\atoms{circ,small}{1/p={-60:0.5}, 2/p={-120:0.5}}
\draw (0)edge[mark={three dots,a}]++(60:0.5) (0)edge[mark={three dots,a}]++(120:0.5) (0)--(1) (1)--++(-60:0.5) (0)--(2) (2)--++(-120:0.5);
\end{tikzpicture}
\;.
\end{equation}
After this, every input and output is adjacent to a sum tensor.

Having deformed the tensor-network diagram, we denote the number of GHZ tensors by $n$, the number of input/output qubits by $n_\partial$, and the number of sum tensors by $\nrzch$.
We denote the $\nrzch\times n$ adjacency matrix between GHZ and sum tensors by $\checkz$.
We denote the $\nrzch\times n_\partial$ adjacency matrix between open (input/output) indices and sum tensors by $\checkz_\partial$, and let $(\checkz,\checkz_\partial)$ denote the stacked $\nrzch\times (n+n_\partial)$ matrix.
The intentional notational overlap with Section~\ref{sec:theory} already hints at the equivalent mathematical structure.

For talking about spacetime local logical gates, we do not only need the tensor-network graph corresponding to the CSS circuit itself, but also the $Z$ decoding (hyper-)graph.
We denote the $n\times m$ adjacency matrix of the $Z$ decoding graph by $\checkx$.
The rows of $\checkx$ are labeled by the GHZ tensors, and the columns are given by a generating set of \emph{$Z$ detectors}.
A $Z$ detector is a length-$n$ binary vector $d$ (with one entry per GHZ tensor) such that $\checkz d=0$, that is, $d\in\ker(B)$, so we have $\checkz\checkx=0$.
For a CSS circuit that is part of a scalable family, the $Z$ detectors are given by all elements of $\ker(B)$ that are locally supported.
That is, their support is contained inside a $r$-ball on the bipartite graph for some small constant $r$.
We also define the $(n+n_\partial)\times m'$ matrix $(\checkx',\checkx_\partial')^T$ whose columns form a basis for $\ker((\checkz,\checkz_\partial))$.
Similarly, the adjacency matrix $\xdecode^T$ of the $X$ decoding graph has columns corresponding to \emph{$X$ detectors}, which are locally supported elements of $\ker(\checkz^T)$, such that $CB=0$.
Taken together, the decoding graphs and the tensor-network graph form a chain complex
\begin{equation}
\zz_2^m\xrightarrow{A} \zz_2^n\xrightarrow{B}\zz_2^o\xrightarrow{C}\zz_2^p\;.
\end{equation}
However, the $X$ decoding graph $C$ does not show up in the spacetime code-space preservation condition.

Now consider the amplitude of the $+1$-postselected CSS circuit (or tensor network) between some input and some output qubit configuration.
Let $a_\partial\in\zz_2^{n_\partial}$ denote the configuration of all inputs and outputs together.
The amplitude for $a_\partial$ is obtained by summing over all configurations of GHZ tensors that satisfy the sum-to-zero constraints given by the sum tensors.
We refer to this sum over configurations as a \emph{homological path integral},
\begin{equation}
\label{eq:path_integral}
\begin{gathered}
Z(a_\partial) \propto \sum_{a\in \zz_2^n: \checkz a=\checkz_\partial a_\partial} 1\\
=
\sum_{[a_0]: \checkz a_0=\checkz_\partial a_\partial} \sum_{x\in \zz_2^{\nrxch}} 1
\propto \sum_{[a_0]: \checkz a_0=\checkz_\partial a_\partial} 1
\;.
\end{gathered}
\end{equation}
In the second line, we have split the sum over all $a$ into a sum over cohomology classes $[a_0]\in \ker(B)/\img(A)$ with a representative $a_0$ that is compatible with $a_\partial$ on the input/output boundary, as well as a sum over all $x\in \zz_2^m$.
The identification between $a$ and the pair $([a_0],x)$ is given by $a=a_0+\checkx x$.
The sum over $x\in\zz_2^m$ yields a normalization factor $2^m$, which we neglect.
So overall, $Z(a_\partial)$ is given by the number of cohomology classes in $\ker(B)/\img(A)$ compatible with $a_\partial$ on the boundary.

The amplitudes $Z(a_\partial)$ can be interpreted as the coefficient vector of a state defined on both the input and output qubits, which we call the \emph{boundary state}.
The boundary state is a code state of a CSS code that we call the \emph{boundary code}.
The binary vectors describing its $Z$ checks are given by all generators of $\img(\checkz_\partial^T \xdecode^T)$.
The $X$ checks are given by all generators of $\img(\checkx_\partial')$.
For more details on how to translate CSS circuits into tensor-network diagrams and spacetime cohomology, we refer the reader to Refs.~\cite{path_integral_qec,Bombin2023,xy_floquet}.

\myparagraph{Spacetime local logical gates}
A spacetime local logical gate is a way to insert diagonal gates into the circuit at different times, satisfying certain conditions described below.
Like in Section~\ref{sec:theory}, we focus on third-order gates consisting of $CCZ$, $CS$, and $T$ gates.
\footnote{We will see in Section~\ref{sec:nondiagonal_nonclifford} that also certain non-diagonal circuit elements such as Hadamard ($H$), $CZ$ measurements, or weak measurements behave like insertions of $CCZ$, $CS$, or $T$ in spacetime.}
The generalization to higher Clifford levels is analogous to Section~\ref{sec:higher_clifford}.
Each pair of qubit and time step in the CSS circuit can be associated with a wire segment in the circuit or tensor-network diagram, and this wire segment can be associated with the adjacent GHZ tensor.
Thus each $CCZ$ gate can be associated with a triple of GHZ tensors, each $CS$ gate with a pair, and each $T$ gate with an individual GHZ tensor.
In our efficient search algorithm, we consider insertions with $CCZ^c$ gates at a subset $N^{(0)}$ of GHZ tensor triples, $CS^c$ gates at a subset $N^{(1)}$ of GHZ pairs, and $T^c$ at a subset $N^{(2)}$ of individual GHZ tensors.
When viewing the $+1$-postselected circuit as a path integral as in Eq.~\eqref{eq:path_integral}, the inserted diagonal gates become phase factors,
\begin{equation}
\label{eq:twisted_path_integral}
\begin{gathered}
Z(a_\partial) = \sum_{a\in \zz_2^n: \checkz a=\checkz_\partial a_\partial} e^{2\pi i S^{n,\mathbf N}_{\mathbf c}(a)}\\
=
\sum_{[a_0]: \checkz a_0=\checkz_\partial a_\partial} \sum_{x\in \zz_2^{\nrxch}} e^{2\pi i S^{n,\mathbf N}_{\mathbf c}(a_0+\checkx x)}\;.
\end{gathered}
\end{equation}
We define the coefficients $\mathbf c$ to give rise to a \emph{spacetime logical gate} if $\mathbf c$ fulfils Eq.~\eqref{eq:cohomology_invariant}, where $A$ denotes the $Z$-decoding graph instead of the $X$-check matrix.
In this case, Eq.~\eqref{eq:twisted_path_integral} simplifies to a sum over cohomology classes,
\begin{equation}
Z(a_\partial) \propto \sum_{[a_0]: \checkz a_0=\checkz_\partial a_\partial} e^{2\pi i S^{n,\mathbf N}_{\mathbf c}(a_0)}\;.
\end{equation}
So calculating the logical effect of a CSS circuit with a spacetime logical gate is particularly simple.

Most importantly, scalable families of CSS circuits with local spacetime logical gates satisfy the following \emph{spacetime fault-tolerance condition}:
Consider a ball of radius $r+\Delta r$ in the circuit/tensor-network diagram, and remove from it a smaller ball of radius $r$.
Then for $\Delta r$ larger than some constant, the boundary state $Z(a_\partial)$ is a product state with respect to the division into the inner boundary (at radius $r$) and the outer boundary (at radius $r+\Delta r$).
The spacetime fault-tolerance condition ensures fault-tolerance with respect to adversarial noise under postselection.
For more discussion and examples for the spacetime fault-tolerance condition in CSS circuits with local spacetime logical gates, we refer the reader to Refs.~\cite{Davydova2025,twisted_double_code,twisted_color_circuits,non_clifford_benchmarking}.

The code-space preservation condition also ensures that the boundary state $Z(a_\partial)$ is a code state of some well-defined boundary code.
This code is no longer an ordinary Pauli stabilizer code but has stabilizers that (1) contain both Pauli and Clifford operators, and (2) do not always commute.
First of all, the $Z$ stabilizers of the CSS circuit given by $\img(\checkz_\partial^T \xdecode^T)$ are still $Z$ stabilizers after adding the spacetime logical gate.
However, the $X$ stabilizers are modified.
The boundary state is now invariant under flipping a subset of qubits and at the same time multiplying with a phase,
\begin{equation}
Z(a_\partial) = e^{2\pi i (S_{\mathbf c}(a+A'(x))-S_{\mathbf c}(a))} Z(a_\partial+A_\partial'(x))\;,
\end{equation}
where $a$ is any bulk GHZ configuration such that $(B,B_\partial)(a,a_\partial)^T=0$, and is defined up to $a\rightarrow a+Ax$.
So we see that the spacetime code-space preservation condition in Eq.~\eqref{eq:cohomology_invariant} ensures that the modified $X$ stabilizers are independent of the choice of $a$ in the bulk.
The modified $X$ stabilizers are a product of the old $X$ stabilizers and a diagonal phase gate.
As we will see in Section~\ref{sec:qudits}, the phase
\begin{equation}
\Delta_x S_{\mathbf c}(a) \coloneqq S_{\mathbf c}(a+A'(x))-S_{\mathbf c}(a)
\end{equation}
is a derivative of a third-order function, which is a second-order function, and thus of the form
\begin{equation}
\Delta_x S_{\mathbf c}(a) =  \frac12 \sum_{i<j} f^{(0)}_{ij} \ovl a_i \ovl a_j + \frac14 \sum_{i} f^{(1)}_i \ovl a_i\;.
\end{equation}
for some coefficients  $f^{(0)}_{ij}\in\zz_2$, $f^{(1)}_i\in\zz_4$.
That is, the modified $X$ stabilizers are equal to the old ones times a diagonal Clifford gate, which is a product of $CZ$ and $S$ gates.

All in all, the mathematical condition for spacetime logical gates of CSS circuits is the same as for ordinary logical gates in CSS codes.
In particular, the former are also given by the kernel of the abelian group homomorphism $\checkx^{\pullb}$.

\subsection{Higher Clifford levels}
\label{sec:higher_clifford}
We have restricted ourselves to diagonal gates in the third level of the Clifford hierarchy in Section~\ref{sec:theory} for concreteness, but the generalization to higher Clifford levels is straightforward.
In general, we can consider an action $S_{\mathbf c}$ of the form
\begin{equation}
\label{eq:any_level_action}
S_{\mathbf c}(a) = \sum_{0\leq l<l_{\max}}\; \sum_{\mathbf i\in N^{(l)}} \frac{1}{2^{l+1}} c^{(l)}_{\mathbf i} \prod_{i\in \mathbf i} \ovl a_i\;,
\end{equation}
where $N^{(l)}$ is a collection of qubit subsets $\mathbf i\subset \{0,\ldots,n-1\}$.
Each term for $\mathbf i\in N^{(l)}$ corresponds to a $|\mathbf i|$-qubit phase-$1/2^{l+1}$ ansatz gate at the level $l+|\mathbf i|$ of the Clifford hierarchy~\cite{Cui2016}.
Accordingly, we say that the order of each above term is $l+|\mathbf i|$, and the order of the function $S_{\mathbf c}$ is the maximal order of one of its terms.
The group $\gphys$ of all considered gates is thus given by
\begin{equation}
\gphys = \bigtimes_{0\leq l<l_{\max}} \zz_{2^{l+1}}^{|N^{(l)}|}\;.
\end{equation}
Note that in Section~\ref{sec:theory}, we have used $l=0,1,2$ ($l_{\max}=3$), and only used subsets with $|\mathbf i|=3-l$ such that each ansatz gate is exactly in the third level.
Also note that each subset $\mathbf i$ should only occur in $N^{(l)}$ for one single $l$, as the ansatz gate for $\mathbf i\in N^{(l)}$ is contained in the ansatz gate for $\mathbf i\in N^{(l')}$ if $l'>l$.

We can still define the pullback $\checkx^{\pullb}$, and compute its structure coefficients using Eq.~\eqref{eq:lift_expansion} analogous to Section~\ref{sec:constructing_coefficients}.
We find
\begin{equation}
\begin{gathered}
S_{\checkx^{\pullb} \delta^{(l)}_{\mathbf i}}(a)
=S_{\delta^{(l)}_{\mathbf i}} (Aa)
=\frac{1}{2^{l+1}} \prod_{i\in\mathbf i} (\ovl{\checkx a})_i\\
\overset{\eqref{eq:z2hom_set_notation}}{=}\frac{1}{2^{l+1}} \prod_{i\in \mathbf i} (\ovl{\sum_{j\in \checkx_i} a_j})\\
\overset{\eqref{eq:lift_expansion}}{=}\sum_{0\leq \lambda \leq l}\frac{(-1)^{\lambda}}{2^{l+1-\lambda}} \sum_{\substack{\{L_i\subset \checkx_i\}_{i\in\mathbf i}:\\|L_i|\geq 1\forall i\\\sum_{i\in\mathbf i} |L_i| =|\mathbf i|+\lambda}}\; \prod_{j\in \bigcup_{i\in\mathbf i} L_i} \ovl a_j\;.
\end{gathered}
\end{equation}
Here, we have used the notation
\begin{equation}
\label{eq:z2hom_set_notation}
A_i = \{j: A_{ij}=1\}\;,
\end{equation}
and $\delta_{\mathbf i}^{(l)}$ denotes the $\mathbf c$ coefficient vector with entry $1$ at $(l,\mathbf i)$ and entry $0$ elsewhere.

We confirm that $S_{\checkx^{\pullb} \mathbf c}$ is a function of the same form as $S_{\mathbf c}$ in Eq.~\eqref{eq:any_level_action}.
Let $M^{(l)}$ be the collection of subsets of $X$ checks $\mathbf j = \bigcup_{i\in \mathbf i} L_i\subset \{0,\ldots,\nrxch-1\}$ that occur in the image of $\checkx^{\pullb}$ with a phase exponent $l$.
Then $\checkx^{\pullb}$ is a group homomorphism of the form
\begin{equation}
\label{eq:general_2homomorphism}
\checkx^{\pullb}: \bigtimes_{0\leq l<l_{\text{max}}} \zz_{2^{l+1}}^{|N^{(l)}|}\rightarrow \bigtimes_{0\leq l<l_{\text{max}}} \zz_{2^{l+1}}^{|M^{(l)}|}\;.
\end{equation}
The terms in $S_{\checkx^{\pullb} \delta^{(l)}_{\mathbf i}}$ have a phase exponent that is smaller or equal to $l$, and an order/Clifford level that is smaller or equal to $l+|\mathbf i|$.
So both the $l_{\max}$ and the maximal Clifford level in $M^{(l)}$ is smaller or equal than in $N^{(l)}$.
There is also a geometric constraint on $M^{(l)}$:
A subset of $X$ checks can only be in $M^{(l)}$ if there is at least one ansatz gate $\mathbf i\in N^{(l)}$ that overlaps with each of these $X$ checks.
Thus, for a locality-preserving logical gate in a qLDPC code family where each qubit participates in a constant number of physical ansatz gates, both the domain and codomain of $\checkx^{\pullb}$ consist of $O(n)$ factors.

We can also still compute the kernel of $\checkx^{\pullb}$ analogous to Section~\ref{sec:2group_linalg}.
The straightforward generalizations of Algorithms~\ref{alg:z248_z2_kernel} and \ref{alg:full_kernel} are given in Algorithms~\ref{alg:general_z2_kernel} and \ref{alg:general_general_kernel}, and the asymptotic runtime with $O(n)$ factors in a dense implementation is still $O(n^3)$.
\begin{figure}
\begin{algorithm}[H]
\caption{Kernel isomorphism $K$ of $X : \prod_{0\leq l<f} \zz_{2^{l+1}}^{n_l}\rightarrow \zz_2^m$}
\label{alg:general_z2_kernel}
\begin{algorithmic}[1]
\Require Binary coefficient matrix $\coeffs{X_0&X_1&\ldots}$ of $X$
\Ensure $f\times f$ block coefficient matrix of $K$
\State Compute a $\zz_2$ staggered kernel isomorphism $Z$ of $X$
\For{$0\leq l<f$}
Compute the $\zz_2$ image completion $W_l$ of $Z_{ll}$
\EndFor
\State Return $K=$
\[
\hspace*{-1em}
\left|
\begin{array}{cc|cc|cc|c|c}
Z_{00}&0&Z_{01}&0&Z_{02}&0&\ldots&Z_{0(f-1)}\\
0&W_1&\ovl Z_{11}&0&\ovl Z_{12}&0&\ldots&\ovl Z_{1(f-1)}\\
0&0&0&\ovl W_2&\ovl Z_{22}&0&\ldots&\ovl Z_{2(f-1)}\\
0&0&0&0&0&\ovl W_3&\ldots&\ovl Z_{3(f-1)}\\
\ldots&\ldots&\ldots&\ldots&\ldots&\ldots&\ldots&\ldots\\
0&0&0&0&0&0&\ldots&\ovl Z_{(f-1)(f-1)}
\end{array}
\right|
\]
\end{algorithmic}
\end{algorithm}
\end{figure}

\begin{figure}
\begin{algorithm}[H]
\caption{Kernel isomorphism $K$ of $X : \prod_{0\leq l<f} \zz_{2^{l+1}}^{n_l}\rightarrow \prod_{0\leq l<g} \zz_{2^{l+1}}^{m_l}$}
\label{alg:general_general_kernel}
\begin{algorithmic}[1]
\Require $g\times f$ block coefficient matrix of $X$
\Ensure $f\times f$ block coefficient matrix of $K$
\State Set $K=\mathbb 1$
\For{$0\leq l<f$}
\State Use Algorithm~\ref{alg:general_z2_kernel} to compute a kernel isomorphism $L$ of
\[\frac{1}{2^l} \mmod_{2^{l+1}} XK\]
\State Set $K=KL$
\EndFor
\State Return $K$
\end{algorithmic}
\end{algorithm}
\end{figure}

\subsection{Arbitrary diagonal gates}
\label{sec:arbitrary_diagonal}
In principle, we do not need to restrict ourselves to the Clifford hierarchy at all.
We can consider arbitrary diagonal ansatz gates,
\begin{equation}
\label{eq:arbitrary_diagonal_action}
S_{\mathbf c}(a) = \sum_{\mathbf i\in N} \mathbf c_{\mathbf i} \prod_{i\in \mathbf i} \ovl a_i\;,
\end{equation}
where $N$ is a collection of qubit subsets, without dependence on a phase exponent $l$.
Each $\mathbf i$ term corresponds to a $|\mathbf i|$-qubit multi-controlled phase gate for an arbitrary phase $\mathbf c_{\mathbf i}\in \rr/\zz$ which may not be in any level of the Clifford hierarchy.
So in this case, we have
\begin{equation}
\mathbf c\in \gphys = (\rr/\zz)^{|N|}\;.
\end{equation}
We can still compute the coefficients of $\checkx^{\pullb}$ analogous to Section~\ref{sec:higher_clifford},
\begin{equation}
\label{eq:coefficients_non_hierarchy}
\begin{gathered}
S_{\checkx^{\pullb} \phi_{\mathbf i}}(a)
=S_{\phi_{\mathbf i}} (Aa)
=\phi \prod_{i\in \mathbf i} (\ovl{\checkx a})_i\\
=\phi \prod_{i\in \mathbf i} (\ovl{\sum_{j\in \checkx_i} a_j})\\
\overset{\eqref{eq:lift_expansion}}{=}\phi \sum_{\substack{0\leq \lambda\\\leq \sum_{i\in \mathbf i} |A_i|-|\mathbf i|}} (-2)^{\lambda} \sum_{\substack{\{L_i\subset \checkx_i\}_{i\in\mathbf i}:\\|L_i|\geq 1\forall i\\\sum_i |L_i| =|\mathbf i|+\lambda}}\; \prod_{j\in \bigcup_{i\in\mathbf i} L_i} \ovl a_j\;.
\end{gathered}
\end{equation}
Here, $\phi_{\mathbf i}$ denotes the $\mathbf c$ vector with phase $\phi\in\rr/\zz$ at the $|\mathbf i|$-qubit location $\mathbf i$ and phase $0$ elsewhere.
We see that $S_{\checkx^{\pullb} \mathbf c}$ is indeed again a function of the same form as $S_{\mathbf c}$ in Eq.~\eqref{eq:arbitrary_diagonal_action}.
If we denote by $M$ the collection of all subsets of $X$ checks that occur in $S_{\checkx^{\pullb} \phi_{\mathbf i}}$, then $\checkx^{\pullb}$ is a homomorphism of the form
\begin{equation}
\checkx^{\pullb}: (\rr/\zz)^{|N|}\rightarrow(\rr/\zz)^{|M|}\;.
\end{equation}
In contrast to the Clifford-hierarchy case, the support $M$ of $\img(\checkx^{\pullb})$ may consist of arbitrarily large subsets of $X$ checks, and is no longer bounded by the maximal size of $N$ plus a maximal phase exponent $l$.
However, there is still the geometric constraint that a subset of $X$ checks can only be in $M$ if there is an ansatz gate whose qubits overlap with each of these $X$ checks.
Thus, for a qLDPC family code with a constant-depth set of ansatz gates, we still have $|M|=O(n)$ and $|N|=O(n)$.

We can also still efficiently compute the kernel of $\checkx^{\pullb}$:
Any homomorphism $h: \rr/\zz\rightarrow\rr/\zz$ is of the form $h(\phi) = c\phi$ for some $c\in\zz$, so the coefficient matrix specifying $\checkx^{\pullb}$ (computed via Eq.~\eqref{eq:coefficients_non_hierarchy}) is an $|M|\times |N|$ integer matrix.
$\ker(\checkx^{\pullb})$ can be found via the \emph{Smith normal form} of the $|M|\times |N|$ coefficient matrix.

We note that Bravyi-Koenig-type arguments~\cite{Bravyi2012a} restrict transversal/locality-preserving gates on qLDPC codes to the Clifford hierarchy.
While these arguments only make statements about the logical action of gates, it seems like the physical realizations of these gates are also restricted to the Clifford hierarchy.
For the on-site case, this has been shown in Refs.~\cite{Anderson2014,Jochym2017}.
So while we can efficiently find arbitrary diagonal non-hierarchy gates, we may restrict ourselves to the Clifford hierarchy without loss of generality.

\subsection{Prime and composite-dimensional qudits}
\label{sec:qudits}
In this section, we discuss the generalization of our methods to qudits of possibly composite dimensions.
\myparagraph{CSS codes for composite-dimensional qudits}
Let $d_i$ be the dimension of the $i$th qudit.
We associate the basis states of the qudit with the elements of the group $\zz_{d_i}$.
The group of all qudit configurations is the cartesian product of all the $\zz_{d_i}$ groups,
\begin{equation}
\label{eq:mixed_qudit_product_group}
G=\prod_{0\leq i<n} \zz_{d_i}\;.
\end{equation}
The $X$ check matrix is given by a homomorphism $\checkx: H\rightarrow G$ for some abelian group
\begin{equation}
H=\prod_{0\leq i<m} \zz_{e_i}\;,
\end{equation}
where $e_i$ are the orders of the stabilizer generators.
There is one $X$ stabilizer $\sigma_X(h)$ for each $h\in H$, acting on a qudit configuration $a\in G$ as
\begin{equation}
\sigma_X(h)\ket{a} = \ket{a+\checkx(h)}\;.
\end{equation}

\myparagraph{$l$th order functions}
In order to define the terms of the phase function $S_{\mathbf c}$ in the qudit case, we need an algebraic generalization of the $l$th order functions defined for qubits in Eq.~\eqref{eq:any_level_action}, see Ref.~\cite{quadratic_tensors}.
We start by defining the \emph{derivative at the $x$th entry} of a $k$-ary function $q: \bigtimes_{0\leq i<k} G_i\rightarrow \rr/\zz$ as
\begin{equation}
\label{eq:kary_ithorder_function_definition}
\begin{gathered}
\partial_x q: G_0\times\ldots G_x\times G_x\times\ldots G_{k-1}\rightarrow \rr/\zz:\\
\partial_x q(g_0,\ldots,g_x,g_x',\ldots,g_{k-1})\\
\coloneqq q(g_0,\ldots,g_x+g_x',\ldots, g_{k-1})\\
-q(g_0,\ldots,g_x,\ldots, g_{k-1})-q(g_0,\ldots,g_x',\ldots, g_{k-1})
\end{gathered}
\end{equation}
Define an \emph{$l+1$th order $k$-ary function} as a function $q$ such that (1) taking $l+1$ times the derivative (at any sequence of entries $x$) yields the zero function, and (2) it yields zero if any of its arguments is zero.
In other words, the derivative of a $l+1$th order $k$-ary function is an $l$th order $k+1$-ary function, and $0$th order functions are zero.
In particular, $1$st order $k$-ary functions are just $k$-linear functions.
We can use Eq.~\eqref{eq:kary_ithorder_function_definition} to expand $l+1$th order functions with a sum in any argument.
For example, in the first argument, we get,
\begin{equation}
\label{eq:kary_ithorder_expansion}
\begin{gathered}
q(\sum_{0\leq i<i_{\max}} g_i,\ldots)\\
= \sum_{\substack{L\subset \{0,\ldots,i_{\max}-1\}\\1\leq |L|\leq l+1}} \partial_0^{|L|-1} q(g_{L_0},\ldots,g_{L_{|L|-1}}, \ldots)\;.
\end{gathered}
\end{equation}
The set of $l+1$th order $k$-ary functions forms an abelian group under addition, which we can identify with an abstract coefficient group.
For $G_i=\zz_{d_i}$, we denote the abstract coefficient group by $\mathcal C_{\mathbf d}^{(l)}$ where $\mathbf d=(d_i)_{0\leq i<k}$, and we write
\begin{equation}
F^{(l)}_{\mathbf d}(c|a_0,\ldots,a_{k-1})
\end{equation}
for the $l+1$th order $k$-ary function with coefficient $c\in \mathcal C_{\mathbf d}^{(l)}$.

For the qubit case where $d_i=2$ for all $i$, we have $\mathcal C_{\mathbf d}^{(l)}= \zz_{2^{l+1}}$, and
\begin{equation}
F^{(l)}_{(2,2,\ldots)}(c|a_0,\ldots,a_{k-1})
= \frac{c}{2^{l+1}} \prod_{0\leq i<k} \ovl a_i\;.
\end{equation}
However, the coefficient groups $\mathcal C_{\mathbf d}^{(l)}$ for qudits of other dimensions do not generalize as straightforwardly as one might expect.
For example, the coefficient group for $2$nd order functions on a single qutrit ($\mathbf d=(3)$ and $l=1$) is given by $\mathcal C_{\mathbf d}^{(l)} = \zz_3\times \zz_3\neq \zz_{3^2}$, which already shows a qualitatively different behavior from the qubit case.

\myparagraph{$l$th order ansatz gates}
We consider diagonal logical gates defined by phase functions $S_{\mathbf c}$ of the form
\begin{equation}
\label{eq:qudit_action}
S_{\mathbf c}(a) = \sum_{0\leq l<l_{\max}} \sum_{\mathbf i \in N^{(l)}} F^{(l)}_{\mathbf d_{\mathbf i}}(c^{(l)}_{\mathbf i}|a_{\mathbf i})
\;,
\end{equation}
where $N^{(l)}$ is a collection of qudit subsets, $\mathbf d_{\mathbf i}=(d_i)_{i\in \mathbf i}$, $a_{\mathbf i} = (a_{\mathbf i_0}, a_{\mathbf i_1}, \ldots)$, and $\mathbf i_x$ denotes the $x$th element of $\mathbf i\subset\{0,\ldots,n-1\}$.
Each term $F^{(l)}_{\mathbf d_{\mathbf i}}$ corresponds to an ansatz gate at the $l+|\mathbf i|$th level of the Clifford hierarchy.
The coefficient group of all considered physical ansatz gates is given by
\begin{equation}
\gphys = \bigtimes_{l} \bigtimes_{\mathbf i \in N^{(l)}} \mathcal C^{(l)}_{\mathbf d_{\mathbf i}}
\;.
\end{equation}
The pullback $\checkx^{\pullb}$ can be computed as
\begin{equation}
\begin{gathered}
S_{\checkx^{\pullb} \gamma^{(l)}_{\mathbf i}}(a)
=S_{\gamma^{(l)}_{\mathbf i}} (Aa)
=F^{(l)}_{\mathbf d_{\mathbf i}}(\gamma|(Aa)_{\mathbf i})\\
=F^{(l)}_{\mathbf d_{\mathbf i}}(\gamma|(\sum_{j\in A_i} a_j)_{i\in \mathbf i})\\
\overset{\eqref{eq:kary_ithorder_expansion}}{=}\sum_{0\leq \lambda \leq l} \sum_{\substack{\{L_i\subset \checkx_i\}_{i\in\mathbf i}:\\|L_i|\geq 1\forall i\\\sum_{i\in\mathbf i} |L_i| =|\mathbf i|+\lambda}} \Big(\prod_{i\in \mathbf i} (\partial_i)^{|L_i|-1} F^{(l)}_{\mathbf d_{\mathbf i}}\Big)(\gamma|a_{L_{\mathbf i}})\;.
\end{gathered}
\end{equation}
Here, $\gamma^{(l)}_{\mathbf i}$ denotes the coefficient vector $\mathbf c$ with value $\gamma\in \mathcal C^{(l)}_{\mathbf d_{\mathbf i}}$ at order $l+1$ and qudit subset $\mathbf i$ and value zero elsewhere, where $\gamma$ is a generator of $\mathcal C^{(l)}_{\mathbf d_{\mathbf i}}$.
$\partial_i$ denotes the derivative of the argument corresponding to the $a_i$ component of the original function $F_{\mathbf d_{\mathbf i}}^{(l)}$.
That is, when we take the derivative $\partial_i$ for the second time, we take it at one of the two arguments resulting from the previous derivative from the $a_i$ argument -- the outcome is the same for either choice.
Further, we have used the notation $a_{\mathbf L_{\mathbf i}} = (a_l)_{l\in \mathbf L_i, i\in \mathbf i} = (a_{(L_{\mathbf i_0})_0}, a_{(L_{\mathbf i_0})_1}, \ldots, a_{(L_{\mathbf i_1})_0},a_{(L_{\mathbf i_1})_1}, \ldots)$.

Since the derivative of an $l+1$th order function is again a $<l+1$th order function, the pullback $S_{\checkx^{\pullb}\mathbf c}$ is again of the same form as $S_{\mathbf c}$ in Eq.~\eqref{eq:qudit_action}.
We can further see that the terms in the pullback $\checkx^{\pullb} \gamma^{(l)}_{\mathbf i}$ are of order less or equal to $l$, and also have Clifford level less or equal to $l+|\mathbf i|$.
Furthermore, each resulting term acts on at most $|\mathbf i|+l$ qudits.

Any finite abelian group $G$ can be decomposed as
\begin{equation}
\label{eq:abelian_group_decomposition}
G\simeq \prod_{p\text{ prime}} \prod_i \zz_{p^{\lambda_i}}\;.
\end{equation}
Further, $\mathcal C^{(l)}_{\mathbf d}=0$ if $\gcd(d_0,d_1,\ldots)=1$.
So $l+1$th order $k$-ary functions can only couple factors in Eq.~\eqref{eq:abelian_group_decomposition} with the same prime $p$.
As a consequence, any qudit CSS code can be decomposed as a disjoint union of independent CSS codes with different underlying primes $p$, and any $l+1$th order gate is a tensor product of independent $l+1$th order gates acting on individual $p$ blocks.
So we can solve the problem of finding logical gates separately for the different $p$ blocks.
Each $p$ block consists of qudits of dimension $d_i=p^{\lambda_i}$, and also all coefficient groups of $k$-ary $l+1$th order functions consist of factors of the form $\zz_{p^\lambda}$.
Finding logical gates thus involves computing the kernel of a homomorphism between groups with $\zz_{p^\lambda}$ factors.
For $p=2$, we have already given efficient filtration algorithms for this in Section~\ref{sec:higher_clifford}.
The generalization from $p=2$ to other primes $p$ is straightforward.

\subsection{Application to finding non-diagonal gates}
\label{sec:nondiagonal_nonclifford}
Even though we have focussed on finding diagonal local (spacetime) logical gates, our method can also be used to find certain non-diagonal gates.
In Section~\ref{sec:spacetime_logical} we have assumed that the terms of the action $S$ in the path integral in Eq.~\eqref{eq:twisted_path_integral} correspond to diagonal gates that have been inserted into the circuit.
However, certain non-diagonal gates can also be represented as a term of the action of the path integral.
As an example, consider a circuit with a single-qubit $X$ measurement.
The projector corresponding to the $+$ outcome (see Eq.~\eqref{eq:tensor_network_translation}) can be simplified to a tensor network with two GHZ tensors,
\begin{equation}
\begin{tikzpicture}
\atoms{circ,small}{0/}
\atoms{circ,small,all}{1/p={0.6,0}}
\draw (0)--(1) (0)edge[ind=$b$]++(90:0.5) (0)edge[ind=$a$]++(-90:0.5);
\end{tikzpicture}
=
\begin{tikzpicture}
\atoms{circ,small,all}{0/, 1/p={0,0.6}}
\draw (1)edge[ind=$b$]++(90:0.5) (0)edge[ind=$a$]++(-90:0.5);
\end{tikzpicture}
=1\forall a,b
\;.
\end{equation}
Now, consider adding the ansatz term $\frac12 c\ovl a\ovl b$ to the action for $c\in \zz_2$.
If $c=0$, the circuit just contains the single-qubit $X$ measurement.
However, if $c=1$, then the $X$ measurement is replaced by a Hadamard gate,
\begin{equation}
e^{2\pi i \frac12\ovl a\ovl b}=
\mpm{1&1\\1&-1}\propto
\begin{tikzpicture}
\node[draw,rectangle](h){$H$};
\draw (h)edge[ind=$b$]++(90:0.5) (h)edge[ind=$a$]++(-90:0.5);
\end{tikzpicture}
\;.
\end{equation}
That is, our method also applies to finding gates that contain (non-diagonal) Hadamard terms.

Similarly, consider a circuit where two qubits experience a single-qubit $X$ measurement at the same time,
\begin{equation}
\begin{tikzpicture}
\atoms{circ,small,all}{0/, 1/p={0,0.6}, 2/p={0.6,0}, 3/p={0.6,0.6}}
\draw (1)edge[ind=$b_0$]++(90:0.5) (0)edge[ind=$a_0$]++(-90:0.5) (3)edge[ind=$b_1$]++(90:0.5) (2)edge[ind=$a_1$]++(-90:0.5);
\end{tikzpicture}
\;,
\end{equation}
and add ansatz terms coupling $a_0$ and $a_1$ with $b_0$ and $b_1$.
Then a combination of 3rd level terms like
\begin{equation}
\frac14 \ovl a_1\ovl b_1 + \frac12 \ovl a_0\ovl b_1 + \frac12 \ovl a_1\ovl b_0
\end{equation}
corresponds to a 2-qubit quantum Fourier transform acting on the two qubits.

As another example, consider a circuit with a single-qubit $X$ measurement and add the ansatz term $\frac14 c\ovl a\ovl b$ for $c\in \zz_4$.
Then $c=1$ corresponds to an operator
\begin{equation}
e^{2\pi i \frac14\ovl a\ovl b}=
\mpm{1&1\\1&i}
=I_+
\;.
\end{equation}
This operator is not a unitary gate that we can execute in the quantum circuit.
However, it can be seen as the $+1$ operator of a quantum instrument.
The $-1$ operator is given by
\begin{equation}
I_-=
\mpm{1&1\\-1&-i}
\;,
\end{equation}
and we indeed find that
\begin{equation}
\label{eq:instrument}
I_+I_+^\dagger + I_- I_-^\dagger \propto \mathbb 1\;.
\end{equation}
Physically, this instrument is a weak measurement followed by a unitary rotation depending on the measurement outcome.
So for $c=0$, we apply a single-qubit $X$ measurement, for $c=1$, we apply the instrument $I$ above, for $c=2$, we apply a Hadamard gate, and for $c=3$, we apply the complex conjugate of $I$.

As a final example, consider a circuit where some qubit experiences an $X$ measurement while some other qubit remains idle at the same time
\begin{equation}
\begin{tikzpicture}
\draw[ind=$a_0$,startind=$a_0$] (-0.6,-0.5)--++(90:1.6);
\atoms{circ,small,all}{0/, 1/p={0,0.6}}
\draw (1)edge[ind=$b_1$]++(90:0.5) (0)edge[ind=$a_1$]++(-90:0.5);
\end{tikzpicture}
\;.
\end{equation}
Now, add an ansatz gate $\frac12 c \ovl a_0\ovl a_1\ovl b_1$.
Then $c=1$ corresponds to an operator
\begin{equation}
I_+=
\begin{pmatrix}
1&1&0&0\\
1&1&0&0\\
0&0&1&1\\
0&0&1&-1\\
\end{pmatrix}
\;.
\end{equation}
This is again not a unitary, but after adding a second operator
\begin{equation}
I_-=
\begin{pmatrix}
1&-1&0&0\\
-1&1&0&0\\
0&0&1&1\\
0&0&1&-1\\
\end{pmatrix}
\;,
\end{equation}
it becomes a valid quantum instrument satisfying Eq.~\eqref{eq:instrument}.
Physically, this instrument means that, controlled on the first qubit, we exchange the $X$ measurement on the second qubit by a Hadamard gate.

\subsection{Faster algorithms using locality or translation invariance}
\label{sec:faster_locality}
In this section, we briefly discuss the possibility to speed up the algorithms for qLDPC codes by making use of their locality/sparsity.
For a qLDPC code family, the parity check matrices are both row and column sparse, that is, there is a constant number of non-zero entries in each row and each column.
The coefficient matrix of $\checkx^{\pullb}$ inherits that property, though the constant number might increase substantially if there is a large number of ansatz gates.
There are some ways in which sparsity can be used to speed up (finite-group) linear algebra algorithms such as computing the kernel.

First of all, working with a sparse-matrix format might by itself already lead to some speedup.
It is, however, not guaranteed that the matrix will stay sparse when computing the RREF.
Switching from the full RREF to a triangular (PLU) decomposition may help preserve sparsity.
It is also unclear whether sparsity will be preserved through the different stages of Algorithm~\ref{alg:full_kernel}.

Another possibility is to not use Gaussian elimination (RREF/PLU) in the first place.
Wiedemann's algorithm~\cite{Wiedemann1986} is a black-box method that finds a single kernel element using $O(n)$ applications of the matrix to some random initial vector.
For a row-sparse matrix, each matrix application needs time $O(n)$, resulting in a $O(n^2)$ cost per kernel vector, that is, per physical logical gate.
If we want to find the full kernel, then runtime is $O(n^2k)$ where $k$ is the number of kernel generators.
Usually, we expect $k=O(n)$ though, so this does not yield a different asymptotic runtime scaling for finding the full kernel.

An additional possibility is to first compute the \emph{local kernel}, that is, for each qubit we compute all elements of $\ker(\checkx^{\pullb})$ that are supported in a radius-$r$ ball around that qubit, where $r$ is some small constant.
As $r$ is constant, this can be done in time $O(n)$.
The remaining $k'$ non-local kernel elements can be found in time $O(n^2k')$ using Wiedemann's algorithm or similar.
While the local kernel elements correspond to physical logical gates with trivial logical action, the $k'$ non-local elements will often correspond to gates with non-trivial logical action.
For many code families and locality-preserving gate locations, $k'$ is much smaller than $n$, so this would represent a true speedup.
Of course, for a hypothetical locality-preserving logical gate that produces a constant rate of logical non-Clifford gates/magic states, we do have $k'=O(n)$.

We would also like to briefly mention that we can restrict ourselves to translation-invariant locality-preserving logical gates on translation-invariant codes in $d$ dimensions.
In this case, the runtime of the algorithm becomes independent of the system size, and only scales with the number of qubits and ansatz gates per unit cell.
\section{Conclusion}
\label{sec:conclusion}
\myparagraph{Summary}
We have studied fault-tolerant finite-depth diagonal non-Clifford gates on a given CSS code.
We have shown that the condition for a diagonal gate described by coefficients $\mathbf c$ to preserve the code space can be rephrased as $\checkx^{\pullb}\mathbf c=0$ for an abelian group homomorphism $\checkx^{\pullb}$.
We can thus find all locality-preserving logical gates composed from a prescribed set of local ansatz gates by computing the kernel of $\checkx^{\pullb}$.
We have shown how this kernel can be computed fast in practice using a filtration method.
The coefficient matrix of $\checkx^{\pullb}$ is of size $O(n)\times O(n)$, where $n$ is the number of qubits.
A naive dense-matrix implementation for finding the kernel has asymptotic runtime $O(n^3)$, with possibility for improvement making use of locality/sparsity for qLDPC code families.
We have focussed on qubit codes with third-level gates in the beginning for concreteness, but have also discussed the generalization to higher Clifford levels, general non-hierarchy diagonal gates, and composite-dimensional qudits.
In particular, we have shown that the same method can also be used to find ``spacetime logical gates'', which are fault-tolerant insertions of diagonal and certain non-diagonal gates into CSS-type syndrome-extraction circuits.
An implementation of the methods is provided in the code repository Ref.~\cite{diagonal_gate_finder}.

\myparagraph{Outlook}
There are several open future directions.
First, it would be good to further speed up the current implementation using some of the ideas in Section~\ref{sec:faster_locality}.

An open question is to which extent the proposed methods generalize to general (non-CSS) stabilizer codes or Clifford syndrome-extraction circuits.
One challenge here is to define a subset of physical non-Clifford gates in a general stabilizer code/circuit that play a role analogous to diagonal gates in CSS codes, and hence can be analyzed efficiently.
Trivial examples can of course be obtained by starting with diagonal gates on a CSS code and performing a non-CSS basis change.
In particular, our methods also apply to finding logical gates that are diagonal in the $X$ basis by symmetry.

Another interesting direction is to optimize the logical gates found.
Our method finds the space of all logical gates, but it would be nice to find the physical implementation of a fixed logical action with the lowest depth, or lowest number of many-qubit gates.
Another direction is to co-optimize finding high-performance codes with logical gates.
These directions may be feasible by implementing semi-efficient algorithms such as greedy search or integer linear programming on top of our method, however, they are unlikely to have guaranteed polynomial runtime.

An interesting theoretical direction is to use the perspective of this paper to analytically describe logical gates in translation-invariant CSS codes.
The image of the $Z$ check matrix of a translation-invariant CSS code forms an ideal of the group ring $\zz_2^n[\zz^d]$, where $d$ is the spatial dimension and $n$ is the number of qubits per unit cell, a perspective first proposed by Haah~\cite{Haah2012}.
The qubit configurations with non-zero amplitude in the code states are in the kernel of the $Z$ check matrix.
Similarly, the image of $\checkx^{\pullb}$ defines an ideal of the group ring $(\zz_2^{n_0}\times \zz_4^{n_1}\times \zz_8^{n_2}\times \ldots)[\zz^d]$, where $n_0$ is the number of ansatz gates with $Z$-like phases per unit cell, $n_1$ is the number of $S$-like phases, $n_2$ is the number of $T$-like phases, and so on.
$\checkx^{\pullb}$ can thus itself be viewed as the $Z$ check matrix of a translation-invariant CSS code defined over qudits of dimension $2^l$.

\myparagraph{Acknowledgements}
I'd like to thank Dave Aasen and Julio Magdalena de la Fuente for discussions.
This work was supported by the U.S. Army Research Laboratory and the U.S. Army Research Office under contract/grant number W911NF2310255, and by the U.S. Department of Energy, Office of Science, National Quantum Information Science Research Centers, and the Co-design Center for Quantum Advantage (C2QA) under contract number DE-SC0012704.

\bibliography{diagonal_logical_gates}

\appendix

\section{Solving 2-group linear equations}
\label{sec:2group_solving}
In this appendix, we show how to solve the linear equation $Xk=b$ for a homomorphism $X$ as in Eq.~\eqref{eq:xhom}, using the filtration method.
This is important for example for finding physical realizations of a specific logical gate.
We start by finding a solution $k_0 = l_0$ to the equation
\begin{equation}
\label{eq:2group_solve_mod2}
m_2 Xl_0= m_2 b
\end{equation}
This solution can be found by taking the solution of the corresponding $\zz_2$-linear equation $Yk_0= b$ and then lifting it.
The full solution space is then given by $\langle k_0+K_0 l_1\rangle$ for different $l_1$, where $K_0$ is the kernel of $X\mmod 2$.
Next, we want to find a solution $k_1$ to
\begin{equation}
\label{eq:2group_solve_mod4}
m_4 Xk_1= m_4 b\;.
\end{equation}
We know that $k_1$ must also be a solution to Eq.~\eqref{eq:2group_solve_mod2} equation, so we can use the ansatz
\begin{equation}
\label{eq:2group_solve_ansatz1}
k_1=k_0+K_0 l_1\;,
\end{equation}
and get
\begin{equation}
\begin{gathered}
m_4 X(k_0+K_0 l_1) = m_4 b\\
\Rightarrow\quad(\frac12 m_4 XK_0)l_1 = \frac12 m_4(b- Xk_0)\;.
\end{gathered}
\end{equation}
This is again a $\zz_2$-linear equation, so we can find a solution for $l_1$ and plug it back into Eq.~\eqref{eq:2group_solve_ansatz1}.
The full solution space for $k_1$ is given by $\langle k_1 + K_1 l_2\rangle$.
Finally, we want to solve the full equation, $Xk_2=b$.
We know that $k_2$ is also a solution to Eq.~\eqref{eq:2group_solve_mod4}, so we can use the ansatz
\begin{equation}
\label{eq:2group_solve_ansatz2}
k_2=k_1+K_1l_2\;,
\end{equation}
and get
\begin{equation}
\begin{gathered}
X(k_1+K_1l_2)- b = 0\\
\Rightarrow\quad
(\frac14 X K_1) l_2 = \frac14(b- Xk_1)\;.
\end{gathered}
\end{equation}
This is again a $\zz_2$-linear equation, and we can pick a solution $l_2$ and plug it into Eq.~\eqref{eq:2group_solve_ansatz2}.

All in all, we need to solve three $\zz_2$-linear equations where the matrices are given by $m_2 X$, $\frac12 m_4 X K_0$, and $\frac14 X K_1$.
Once we have pre-computed the row-reduced echelon forms of these matrices in time $O(n^3)$, each solution only takes time $O(n^2)$.

The algorithm is summarized in Algorithm~\ref{alg:z248_solve}.
\begin{figure}
\begin{algorithm}[H]
\caption{Solve $Xk=b$ for $X : \zz_2^{n_0} \times \zz_4^{n_1} \times \zz_8^{n_2} \to \zz_2^{m_0}\times \zz_4^{m_1}\times \zz_8^{m_2}$}
\label{alg:z248_solve}
\begin{algorithmic}[1]
\Require $3\times 3$ block coefficient matrix of $X$, coefficient matrices $K_0$, $K_1$ from Algorithm~\ref{alg:z248_z2_kernel}
\Ensure $3$-block vector of a solution $k$
\State Find a solution $l_0=k_0$ to
\[(m_2 X) l_0 = m_2 b\]
\State Find a solution $l_1$ to
\[(\frac12 m_4 XK_0)l_1 = \frac12 m_4(b- Xk_0)\]
\State Compute $k_1=k_0+K_0l_1$
\State Find a solution $l_2$ to
\[(\frac14 X K_1) l_2 = \frac14(b- Xk_1)\]
\State Return $k=k_2=k_1+K_1l_2$
\end{algorithmic}
\end{algorithm}
\end{figure}

\bibliographystyle{quantum}

\end{document}